\documentclass[runningheads]{llncs}

\usepackage{amsfonts}
\usepackage{amsmath}
\usepackage{booktabs}
\usepackage{graphicx}
\usepackage{listings}
\usepackage{multicol}
\usepackage{pdflscape}
\usepackage{subcaption}
\usepackage{stmaryrd}
\usepackage{tikz}
\usepackage{url}
\usepackage{xcolor}
\usetikzlibrary{quantikz2}

\usepackage[capitalise]{cleveref}
\crefname{algorithm}{Alg.}{Algs.}
\crefname{section}{Sec.}{Secs.}
\crefname{definition}{Def.}{Defs.}
\crefname{table}{Tab.}{Tabs.}
\crefname{example}{Ex.}{Exs.}
\crefname{proposition}{Prop.}{Props.}
\crefname{theorem}{Thm.}{Thms.}
\crefname{corollary}{Cor.}{Cors.}
\crefname{appendix}{Appx.}{Appxs.}

\newcommand{\diag}{\operatorname{diag}}
\newcommand{\sembrack}[1]{\llbracket #1 \rrbracket}

\newboolean{isExtended}
\setboolean{isExtended}{true}
\newcommand{\appendixcite}[1]{%
  \ifthenelse{\boolean{isExtended}}%
             {\cref{#1}}%
             {the extended version}}
\newcommand{\addappendix}[1]%
  {\ifthenelse{\boolean{isExtended}}{#1}{}}

\newboolean{isAnonymous}
\setboolean{isAnonymous}{false}
\newcommand{\anonymize}[1]%
  {\ifthenelse{\boolean{isAnonymous}}{}{#1}}
\newcommand{\redact}[1]%
  {\ifthenelse{\boolean{isAnonymous}}{\textbf{omitted for anonymity}}{#1}}

\begin{document}

\lstdefinestyle{qasm}{
    language=c++,
    keywordstyle=\color{blue},
    morekeywords={pow, ctrl, gate, gphase, angle, inv, negctrl, OPENQASM, include, reset, qubit, measure},
    extendedchars=true,
    upquote=true,
    basicstyle=\scriptsize\ttfamily,
    columns=[c]fixed,
    aboveskip=0mm,
    belowskip=2mm,
    keepspaces=true,
    mathescape=true,
    showstringspaces=false,
    showspaces=false,
    numbers=left,
    numberstyle=\tiny,
    numbersep=4pt,
    tabsize=2,
    breaklines=true,
    showtabs=false,
    captionpos=b,
    escapechar=¤,
	comment=[l]{//},
	morecomment=[s]{/*}{*/},
	commentstyle=\color{gray}\ttfamily,
    xleftmargin=1.3em,%
}

\lstdefinestyle{quipper}{
    keywordstyle=\color{blue},
    morekeywords={QGate, GPhase, QRot, with, controls},
    extendedchars=true,
    upquote=true,
    basicstyle=\scriptsize\ttfamily,
    columns=[c]fixed,
    aboveskip=0mm,
    belowskip=2mm,
    keepspaces=true,
    mathescape=true,
    showstringspaces=false,
    showspaces=false,
    numbers=left,
    numberstyle=\scriptsize,
    numbersep=4pt,
    tabsize=2,
    breaklines=true,
    showtabs=false,
    captionpos=b,
    escapechar=¤,
	comment=[l]{//},
	morecomment=[s]{/*}{*/},
	commentstyle=\color{gray}\ttfamily,
    xleftmargin=1.3em,%
}

\lstdefinestyle{haskell}{
    keywordstyle=\color{blue},
    extendedchars=true,
    upquote=true,
    basicstyle=\scriptsize\ttfamily,
    columns=[c]fixed,
    aboveskip=0mm,
    belowskip=2mm,
    keepspaces=true,
    mathescape=true,
    showstringspaces=false,
    showspaces=false,
    numbers=left,
    numberstyle=\scriptsize,
    numbersep=4pt,
    tabsize=2,
    breaklines=true,
    showtabs=false,
    captionpos=b,
    escapechar=¤,
	comment=[l]{//},
	morecomment=[s]{/*}{*/},
	commentstyle=\color{gray}\ttfamily,
    xleftmargin=1.3em,%
}

\ifthenelse{\boolean{isExtended}}%
  {\title{LinguaQuanta: Towards a Quantum Transpiler Between OpenQASM and Quipper (Extended)}}%
  {\title{LinguaQuanta: Towards a Quantum Transpiler Between OpenQASM and Quipper}}
\titlerunning{LinguaQuanta: Towards a Quantum Transpiler}
\anonymize{\author{Scott Wesley}}
\anonymize{\authorrunning{S. Wesley}}
\anonymize{\institute{Dalhousie University, Halifax NS, Canada}}
\maketitle
\begin{abstract}
    As quantum computing evolves, many important questions emerge, such as how best to represent quantum programs, and how to promote interoperability between quantum program analysis tools.
    These questions arise naturally in the design of quantum transpilers, which translate between quantum programming languages.
    In this paper, we take a step towards answering these questions by identifying challenges and best practices in quantum transpiler design.
    We base these recommendations on our experience designing \emph{LinguaQuanta}, a quantum transpiler between Quipper and OpenQASM.
    First, we provide categorical specifications for quantum transpilers, which aim to encapsulate the core principles of the UNIX philosophy.
    We then identify quantum circuit decompositions which we expect to be useful in quantum transpilation.
    With these foundations in place, we then discuss challenges faced during the implementation of LinguaQuanta, such as ancilla management and \emph{stability under round translation}.
    To show that LinguaQuanta works in practice, a short tutorial is given for the example of quantum phase estimation.
    We conclude with recommendations for the future of LinguaQuanta, and for quantum software development tools more broadly.
\end{abstract}

\section{Introduction}

In 2018, a perspective article in Quantum Views proclaimed the dawn of quantum computing~\cite{Ross2018}.
At this time IBM had released a small, but universal, quantum computer~\cite{Mandelbaum2021} which reignited interest in quantum programming.
Over the five years following this landmark development, researchers have developed considerably faster \emph{superconducting} quantum computers with physical storage exceeding $1000$ qubits~\cite{Choi2023}.
In turn, the landscape of quantum programming languages has also exploded, with many new languages emerging~(e.g.,~\cite{BichselBaader2020,Circ2023,CrossJavadiAbhari2022,GreenLumsdaine2013a,Azure2017}) each with their own compilers and development tools~(e.g.,~\cite{McClean2020,RossSelinger2016,WatkinsNguyen2023}).

As we move forward from the dawn of quantum computing, we are now confronted with many important questions, such as how to represent quantum programs, and how best to promote interoperability between quantum program analysis tools.
In classical computing, this problem was solved by intermediate representation (IR) languages such as LLVM-IR~\cite{LattnerAdve2004}, and by standard logical formulations such as constrained Horn clauses~\cite{GrebenshchikovLopes2012}.
In this paper, we focus on the first approach, and consider IR languages for quantum computing.
One popular contender in this area is the OpenQASM assembly language~\cite{CrossBishop2017,CrossJavadiAbhari2022}.

For OpenQASM to act as an IR language, we must first develop tools to perform source-to-source translation from popular quantum programming languages, to the OpenQASM language.
Such source-to-source translators are known as \emph{transpilers}.
To date, there has been very little focus on quantum transpilation (outside of circuit optimization), so best practices in this area are poorly defined.
However, prior work, such as~\cite{Leblond2023}, highlights the need for quantum transpilation in the design and prototyping of new compilation tools.
In this paper we set out to identify best practices and challenges for quantum transpilation, through practical experience with a concrete example.

To this end, we propose \emph{LinguaQuanta}, a quantum transpiler between the functional Quipper language~\cite{GreenLumsdaine2013a}, and the imperative OpenQASM language.
Focus is given to OpenQASM 3~\cite{CrossJavadiAbhari2022}, but legacy support is provided for the 2.0 standard~\cite{CrossBishop2017}. 
\cref{Sect:Background} reviews quantum computation.
\cref{Sect:Pipeline} provides categorical specifications for an abstract quantum transpiler, as informed by best practices in software development.
These specifications prioritize small transpilation tools built using compositional design principles.
\cref{Sect:Decompositions} identifies standard quantum circuit decomposition techniques which prove useful in LinguaQuanta.
\cref{Sect:Challenges} identifies challenges faced during the design of LinguaQuanta, such as reconciling different measurement abstractions, and circumventing the lack of ancillas in OpenQASM.
\cref{Sect:Tutorial} provides an overview of LinguaQuanta, using the quantum phase estimation algorithm as an example.
\cref{Sect:Discussion} discusses lessons learned from the development of LinguaQuanta, and plots a course for future research.
Appendices can be found in the extended version of the paper~\cite{Wesley2024}.

\section{A Brief Introduction to Quantum Computation}
\label{Sect:Background}

This section provides a brief introduction to quantum computation.
For a more comprehensive introduction, we refer the reader to~\cite{NielsenChuang2011}.
We assume the reader is familiar with linear algebra.
If not, we refer the reader to an introductory text such as~\cite{Axler2014}.
A \emph{quantum bit (qubit)} can be in a state of $0$ or $1$, similar to classical computation.
We denote these basis states $\ket{0}$ and $\ket{1}$ respectively, and represent these states by the two-dimensional vectors $\ket{0} = \begin{bmatrix} 1 & 0 \end{bmatrix}^T$ and $\ket{1} = \begin{bmatrix} 0 & 1 \end{bmatrix}^T$.
A qubit $\ket{\psi}$ can also exist in a \emph{superposition} of both $\ket{0}$ and $\ket{1}$~\cite{NielsenChuang2011}.
Formally, $\ket{\psi} = \alpha \ket{0} + \beta \ket{0}$ for $\alpha$ and $\beta$ complex numbers satisfying $|\alpha|^2 + |\beta|^2 = 1$.

An $n$-qubit quantum system has $2^n$ possible basis states, corresponding to the $2^n$ binary strings of length $n$.
As before, the system may also be in a superposition of basis states.
The squared norms of these coefficients must still sum to $1$.
As an example, the basis states for a $2$-qubit quantum system are as follows.
{\begin{align*}
    \ket{00} &= \begin{bmatrix}
        1 & 0 & 0 & 0
    \end{bmatrix}^T
    &
    \ket{01} &= \begin{bmatrix}
        0 & 1 & 0 & 0
    \end{bmatrix}^T
    &
    \ket{10} &= \begin{bmatrix}
        0 & 0 & 1 & 0
    \end{bmatrix}^T
    &
    \ket{11} &= \begin{bmatrix}
        0 & 0 & 0 & 1
    \end{bmatrix}^T
\end{align*}}%
An arbitrary $2$-qubit state is a linear combination $\alpha \ket{00} + \beta \ket{01} + \gamma \ket{10} + \rho \ket{11}$ satisfying $|\alpha|^2 + |\beta|^2 + |\gamma|^2 + |\rho|^2 = 1$.

In quantum mechanics, the measurement of a qubit $\ket{\psi}$ collapses $\ket{\psi}$ to either state $\ket{0}$ or state $\ket{1}$.
This means that if $\ket{\psi} = \alpha \ket{0} + \beta \ket{1}$, it is impossible to determine the values of $\alpha$ and $\beta$.
However, it is known that the probability $\ket{\psi}$ collapses to $\ket{0}$ is $|\alpha|^2$ and the probability that $\ket{\psi}$ collapses to $\ket{1}$ is $|\beta|^2$, where as before $|\alpha|^2 + |\beta|^2 = 1$ \cite{NielsenChuang2011}.
More generally, if $\ket{\psi}$ can be written as $\alpha \ket{\psi_1}\ket{0} + \beta \ket{\psi_2}\ket{1}$, then the probability of measuring the last qubit in state $\ket{0}$ is $|\alpha|^2$ and the probability of measuring the last qubit in state $\ket{1}$ is $|\beta|^2$.

\subsection{Unitary Operations on Quantum Systems}

Quantum systems evolve according to unitary dynamics.
In quantum computation discrete unitary operations are used to encode computational problems.
Before defining unitary operations, some results from linear algebra must be recalled~(see, e.g.~\cite{Axler2014}).
The \emph{adjoint} of an $n \times n$ complex matrix $M$ is the conjugate transpose of $M$, denoted $M^{\dagger} = \overline{M}{}^{T}$.
An $n \times n$ matrix $M$ is said to be unitary if $MM^{\dagger} = M^{\dagger}M = I$, that is, $M^{-1} = M^{\dagger}$.
For example, both rotations and global phases are examples of unitary operations (see~\appendixcite{Appendix:RotBG}).
The $n \times n$ unitary matrices form a group under multiplication denoted $\mathcal{U}(n)$.

A quantum operation on an $n$-qubit quantum system is a linear, invertible operator that sends quantum states to quantum states.
It turns out that the $n$-qubit quantum operators are precisely the matrices in $\mathcal{U}(2^n)$~\cite{NielsenChuang2011}.
For example, there exists an operator $X \in \mathcal{U}(2)$ such that $X \ket{0} = \ket{1}$ and $X \ket{1} = \ket{0}$.
It is not hard to see that $X$ is the quantum generalization of a classical not-gate.
It follows from this analogy that $X^{\dagger} = X$ since $X^{\dagger}$ is the inverse to $X$.

In reversible classical computation, another important gate is the controlled not-gate, denoted $C(X)$~\cite{NielsenChuang2011}.
The $C(X)$ gate belongs to $\mathcal{U}(4)$ and negates the second bit if and only if the first bit is set to $1$.
That is to say $C(X) \ket{00} = \ket{00}$, $C(X) \ket{01} = \ket{01}$, $C(X) \ket{10} = \ket{11}$, and $C(X) \ket{11} = \ket{10}$.
It follows that $C(X)$ is also unitary.
The matrix $C(X)$ is the \emph{controlled} version of $X$.
It turns out that every operator $G \in \mathcal{U}(2^n)$ has a controlled version $C(G) \in \mathcal{U}(2^{n+1})$ defined by $C(G) = \begin{bmatrix} I & 0 \\ 0 & G \end{bmatrix}$, where $I$ is the $(2^n) \times (2^n)$ identity matrix.

\subsection{Compositionality of Quantum Systems}

Quantum operators can be composed in sequence and in parallel, just as gates can be composed in classical computation.
The case of sequential composition is the most straightforward.
Given operators $M \in \mathcal{U}(2^n)$ and $N \in \mathcal{U}(2^n)$, there exists an operator $N \circ M \in \mathcal{U}(2^n)$ such that applying $N \circ M$ is equivalent to first applying $M$ and then applying $N$.
This sequential composition corresponds to matrix multiplication.
It follows almost immediately that $(N \circ M)^\dagger = M^\dagger \circ N^\dagger$.

Parallel composition is more complicated.
To understand this challenge, it is sufficient to look at the composition of two quantum systems, each with a single qubit.
Before composition, each system has two basis states: $\ket{0}$ and $\ket{1}$.
After composition, every combination of the two qubits must be considered.
This yields four states: $\ket{00}$, $\ket{01}$, $\ket{10}$, and $\ket{11}$.
In general, the composition of an $n$ state system with an $m$ state system yields an $n \cdot m$ state system consisting of all pairwise combinations of states~\cite{NielsenChuang2011}.
In the case of qubits, an $n$ qubit system has $2^n$ states, an $m$ qubit system has $2^m$ states, and their parallel composition has $2^{n+m}$ states.
This parallel composition, denoted $\otimes$, is known as the Kronecker tensor product and extends to matrix composition in the following way~\cite{NielsenChuang2011}.
\begin{equation*}
    \begin{bmatrix}
        c_{1,1} & c_{1,2} & \cdots & c_{1,n} \\
        c_{2,1} & c_{2,2} & \cdots & c_{2,n} \\
        \vdots & \ddots & \vdots & \vdots \\
        c_{m,1} & c_{m,2} & \cdots & c_{m,n}
    \end{bmatrix}
    \otimes
    M
    =
    \begin{bmatrix}
        c_{1,1} M & c_{1,2} M & \cdots & c_{1,n} M \\
        c_{2,1} M & c_{2,2} M & \cdots & c_{2,n} M \\
        \vdots & \ddots & \vdots & \vdots \\
        c_{m,1} M & c_{m,2} M & \cdots & c_{m,n} M
    \end{bmatrix}
\end{equation*}
It follows that $(M \otimes N)(\ket{\psi} \otimes \ket{\varphi}) = (M\ket{\phi}) \otimes (N\ket{\varphi})$ as desired.
By convention $\ket{j} \otimes \ket{k}$ is the concatenation $\ket{jk}$ of $j$ with $k$.
For example, $\ket{0} \otimes \ket{11} = \ket{011}$.

\subsection{Circuit Semantics and Compositionality}
\label{Sect:Background:String}

The formalism discussed so far admits a graphical calculus corresponding to string diagrams in a monoidal category~\cite{Selinger2010}.
We use these diagrams freely.
However, the resulting diagrams are similar to classical circuits, and it suffices that the reader only knows how to interpret string diagrams as matrices.
We do not assume any familiarity with category theory, but we do note that it forms a great basis for compositional reasoning with diagrammatic calculi (e.g.,~\cite{KissingerZamdzhiev2015,MacLane2010,Selinger2010}).

In quantum circuit diagrams, horizontal wires represents program state.
A wire consisting of a single horizontal line indicates a single qubit.
A wire consisting of two horizontal lines indicates a single classical bit.
If $\ket{\varphi}$ is written to the left of a wire, then the wire is initialized in state $\ket{\varphi}$.
Likewise, if $\ket{\varphi}$ is written to the right of a wire, then the wire is asserted to be in state $\ket{\varphi}$, before the wire is discarded.
A wire with a state on both the left and the right represents a temporary variable, and is referred to as an \emph{ancilla} in quantum computation.
For example,
\scalebox{0.7}{\begin{quantikz}[wire types={c}]
        \lstick{\ket{0}} & & \rstick{\ket{0}}
\end{quantikz}}
depicts a classical ancilla initialized in state $\ket{0}$, and
\scalebox{0.7}{\begin{quantikz}
    \lstick{\ket{1}} & & \rstick{\ket{1}}
\end{quantikz}}
depicts a quantum ancilla initialized in state $\ket{1}$.

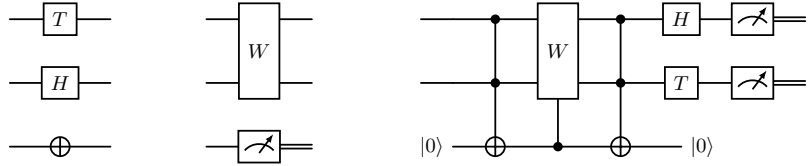
\begin{figure}[t]
    \begin{subfigure}[b]{0.48\textwidth}
    \begin{center}
    \scalebox{0.85}{\begin{quantikz}
        & \gate{T} & & \setwiretype{n} & \ghost{H} & \ghost{H} & \gate[2]{W} \setwiretype{q} & \\ 
        & \gate{H} & & \setwiretype{n} & \ghost{H} & \ghost{H} & \setwiretype{q} & \\
        & \targ{} & & \setwiretype{n} & & & \meter{} \setwiretype{q} &\setwiretype{c}
    \end{quantikz}}
    \end{center}
    \caption{Unitary and measurement gates.}
    \label{Fig:GateEx}
    \end{subfigure}
    \begin{subfigure}[b]{0.48\textwidth}
    \begin{center}
    \scalebox{0.85}{\begin{quantikz}[wire types={q,q,n}]
        & & \ctrl{1} & \gate[2]{W} & \ctrl{1} & \gate{H} & \meter{} &\setwiretype{c} \\
        & & \ctrl{1} & & \ctrl{1} & \gate{T} & \meter{} &\setwiretype{c} \\
        \ghost{H} &\lstick{\ket{0}} & \targ{}\setwiretype{q} & \ctrl{-1} & \targ{} & \rstick{\ket{0}} & \setwiretype{n}
    \end{quantikz}}
    \end{center}
    \caption{A simple string diagram.}
    \label{Fig:CircEx}
    \end{subfigure}
    \caption{A collection of quantum gates, and a circuit constructed from those gates.}
\end{figure}

Quantum operators are depicted as boxes.
The qubits acted on by the operator are depicted as wires passing through the box.
Since each operator is unitary, the number of wires entering and leaving each box are always equal.
Two exceptions to this rule are the not-gate, which is often depicted as an $\oplus$, and the measurement gate, which is often depicted as a gauge.
See \cref{Fig:GateEx} for examples of the $T$ and $H$ gates (each acting on one qubit), the $W$ gate (acting on two qubits), the $X$ gate (acting on one qubit), and a single qubit measurement.

Circuits are then constructed from the parallel and sequential composition of these basic building blocks.
Sequential composition is depicted by joining wires end-to-end.
Parallel composition is depicted by running wires parallel to one-another.
Additionally, controls are depicted by running a vertical wire from a dot (on each control qubit) to the controlled gate.
See \cref{Fig:CircEx} for a circuit making use of a controlled $W$ gate and two doubly-controlled $X$ gates.

\subsection{An Introduction to OpenQASM and Quipper}
\label{Sect:Background:Lang}

\begin{figure}[t]
    \centering
    \setlength{\columnsep}{-2.3cm}
\begin{lstlisting}[style=qasm,
                     multicols=2,
                     basicstyle=\tiny\ttfamily]
OPENQASM 3;
include "stdgates.inc";

qubit phi;
qubit[3] x;
reset x[0]; reset x[1]; reset x[2];

h x[0]; h x[1]; h x[2];

pow(pow(2,2)) @ ctrl @ t x[0], phi; 
pow(pow(2,1)) @ ctrl @ t x[1], phi;
ctrl @ t x[2], phi;

h x[0]; cp(pi / 2) x[1], x[0]; cp(pi / 4) x[2], x[0];
h x[1]; cp(pi / 2) x[2], x[1];
h x[2];

measure x[0]; measure x[1]; measure x[2];
\end{lstlisting}
    \caption{An implementation of QPE in OpenQASM 3 with $3$ digits of accuracy.}
    \label{Fig:QasmQPE:Orig}
    \setlength{\columnsep}{0cm}
\end{figure}

\begin{figure}[t]
    \centering
    \includegraphics[scale=1.1]{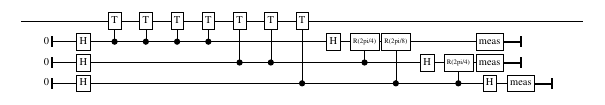}
    \caption{An implementation of QPE in Quipper with $3$ digits of accuracy.}
    \label{Fig:QuipQPE:Orig}
\end{figure}

Two popular languages for quantum algorithm design are OpenQASM 3~\cite{CrossJavadiAbhari2022} and Quipper~\cite{GreenLumsdaine2013a}.
Both languages take a circuit-based approach to quantum computation.
In this model, a program first declares a finite set of qubits, and optionally, a finite set of classical bits (\emph{cbits}).
In OpenQASM, the set of qubits is immutable, whereas in Quipper, the set of qubits can expand or shrink.
Using these declarations, a program then specifies a sequence of unitary gates, classical gates, and measurement gates, which are applied to the qubits and cbits in order.
In OpenQASM 3, these operations are specified in an imperative style, whereas in Quipper, these operations are declared in a functional style\footnote{Quipper also offers an imperative interface to these functional primitives.}.
The functional programs written in Quipper are categorically equivalent to the circuit diagrams outlined in \cref{Sect:Background:String}.
For ease of presentation, this correspondence is used to depict all Quipper programs as circuit diagrams.
In contrast, OpenQASM 3 programs are presented syntactically, as is conventional for imperative languages.
Note that this section does not provide a comprehensive introduction to the features of either language.
Instead, this section only highlights the language features which are relevant to the design of LinguaQuanta.

For concreteness, we consider the \emph{Quantum Phase Estimation (QPE)} algorithm, which uses quantum computation to approximate the eigenvalues of a predetermined operator $U$.
We take $U$ to be the diagonal gate $T$, which has entries $1$ and $e^{i\pi/4}$
Standard implementations of the QPE algorithm, for both OpenQASM 3 and Quipper, can be found in \cref{Fig:QasmQPE:Orig} and \cref{Fig:QuipQPE:Orig} respectively.

The first major distinction between OpenQASM and Quipper is the way in which qubits are modelled.
In OpenQASM, qubits are variables that are referenced by name.
For example, the statement \lstinline[style=qasm]{qubit phi;} on line~4 of \cref{Fig:QasmQPE:Orig} declares a new qubit bound to the identifier \lstinline[style=qasm]{phi}.
In contrast, Quipper models qubits as wires in circuit diagrams.
Each wire in the circuit diagram is assigned a unique index, which is used to reference the corresponding qubit throughout the program.
By default, these indices are consecutive starting from zero.
For example, the uppermost wire in \cref{Fig:QuipQPE:Orig} has index $0$ and corresponds to the variable \lstinline[style=qasm]{phi} in \cref{Fig:QasmQPE:Orig}.
However, Quipper also supports ancilla initialization, ancilla termination, qubit preparation (casting a cbit to a qubit), and adjoint qubit preparation (casting a qubit in a pure state to a cbit), which can yield non-consecutive indices.
In most Quipper programs, only ancilla initialization and ancilla termination are used.
These operations are indicated by vertical bars, as illustrated by the bottom three wires in \cref{Fig:QuipQPE:Orig}.
It should be noted that OpenQASM does not support ancilla management nor qubit preparation, and consequently, the number of qubits in an OpenQASM program is static across program execution.

In terms of unitary operators, OpenQASM and Quipper are quite similar.
Both languages allow users to define their own unitary operators, though only OpenQASM allows for the semantics to be declared alongside the syntactic definition.
Often, these interfaces are unnecessary, since both OpenQASM and Quipper provide standard libraries for unitary operators (in OpenQASM, these gates are imported explicitly using an \lstinline{include} statement, as illustrated on line~2 of \cref{Fig:QasmQPE:Orig}).
For example, \lstinline[style=qasm]{cx c, t;} is an OpenQASM instruction to apply the standard library controlled not-gate to target qubit $t$ with control qubit $c$.
In Quipper, this same instruction would be represented by an arity-$2$ gate acting on wires $c$ and $t$.
A full list of standard operators can be found in \appendixcite{Appendix:Decomps}.
Of note are the rotational gates provided by both languages.
A rotational gate $R( \theta )$ is a family of unitary operators defined by a parameter $\theta$.
In Quipper, rotational gates are restricted to a single parameter, whereas OpenQASM $3$ gates allow for arbitrarily many parameters.
For example, OpenQASM $3$ provides the $\mathcal{U}( \theta, \rho, \lambda )$ gate which can be used to implement any single-qubit unitary operator up to global phase~(i.e., multiplication by a complex scalar of norm $1$).
In the case of \cref{Fig:QasmQPE:Quip}, the rotation operator \lstinline[style=qasm]{cp(pi / 2) x[2], x[1]} on line~15 implements a controlled rotation by a global phase of $\pi / 2$.

Both OpenQASM $3$ and Quipper allow for unitary operators to be controlled and inverted.
Additionally, OpenQASM $3$ allows for unitary operators to be exponentiated.
In OpenQASM $3$, these transformations are indicated using \emph{modifier} annotations.
An example of a control modifier appears in the statement \lstinline[style=qasm]{ctrl @ t x[2], phi} on line~12 of \cref{Fig:QasmQPE:Orig}, which applies $C(T)$ to $\text{\lstinline[style=qasm]{x[2]}} \otimes \text{\lstinline[style=qasm]{phi}}$.
Likewise, an example of an \emph{inverse modifier} would be \lstinline[style=qasm]{inv @ t phi}, which applies $T^{\dagger}$ to qubit \lstinline[style=qasm]{phi}.
Finally, an example of a \emph{power modifier} appears in the statement \lstinline[style=qasm]{pow(2) @ ctrl @ t x[1], phi} on line~11 of \cref{Fig:QasmQPE:Orig} which applies $C(U)^2$ to $\text{\lstinline[style=qasm]{x[1]}} \otimes \text{\lstinline[style=qasm]{phi}}$.
The control modifiers and inverse modifiers in Quipper are depicted graphically as described for circuit diagrams.

Finally, both OpenQASM and Quipper provide primitives for measurement.
The semantics of these primitives reflect the fact that OpenQASM is imperative, whereas Quipper is functional.
The \lstinline[style=qasm]{bit c = measure q;} statement in OpenQASM can be understood as applying an impure function \lstinline[style=qasm]{measure} to qubit $q$ with output cbit $c$.
In this statement, $c$ is the measurement result, and $q$ collapses according to $c$ as a side-effect of \lstinline[style=qasm]{measure}.
This is used on line~18 of \cref{Fig:QuipQPE:Orig}, to obtain the first three bits of the eigenvalue approximation.
Conversely, the measure gate in Quipper is a pure function, which takes as input a qubit $q$, and returns as output a cbit $c$ which corresponds to the collapsed state of $q$.
In \cref{Fig:QuipQPE:Orig}, this corresponds to the three \lstinline[style=qasm]{meas} boxes which convert wires from qubits to cbits.

\section{Pipeline Designs and Categorical Specifications}
\label{Sect:Pipeline}

The \emph{UNIX Philosophy} is a design philosophy which aims to develop software that is simple, general, intelligible, and above all else, effective for programming researchers~\cite{McIlroyPinson1978}.
In~\cite{Salus1994}, the UNIX Philosophy was summarized as follows: (1) write programs that do one thing and do it well; (2) write programs that work together; (3) write programs to handle text streams.
Since transpilation is a classical task, we have followed the UNIX philosophy and designed LinguaQuanta as a collection of simple utilities which compose together to achieve various modes of transpilation.
These components are as follows.
\begin{enumerate}
    \item \textsc{ElimCtrls}: A tool to inline control modifiers in an OpenQASM program.
    \item \textsc{ElimInvs}: A tool to inline inversion modifiers in an OpenQASM program.
    \item \textsc{ElimPows}: A tool to inline power modifiers in an OpenQASM program.
    \item \textsc{ElimFuns}: A tool to inline built-in OpenQASM functions.
    \item \textsc{QasmToQuip}: A tool to transpile from OpenQASM to Quipper.
    \item \textsc{QuipToQasm}: A tool to transpile from Quipper to OpenQASM.
    \item \textsc{RegMerge}: A tool to combine all registers in an OpenQASM program.
    \item \textsc{ToLsc}: A tool to transpile OpenQASM 2.0 to the subset described in~\cite{WatkinsNguyen2023}.
    \item \textsc{ToQasm2}: A tool to convert between OpenQASM 2.0 and 3.
\end{enumerate}
Details on how tools compose can be found on the LinguaQuanta GitHub page\footnote{\redact{\url{https://github.com/onestruggler/qasm-quipper}}}.

The composition of tools in LinguaQuanta can be understood more abstractly as the composition of functions mapping source languages to target languages.
Then, it is natural to ask what categorical properties should tool composition satisfy.
For concreteness, consider the functions $L_1 \xrightarrow{T_1} L_2 \xrightarrow{T_2} L_1$ where $T_1$ denotes \textsc{QasmToQuip} and $T_2$ denotes \textsc{QuipToQasm}.
Then $T_2 \circ T_1$ is a round translation starting from OpenQASM.
It is tempting to require that $T_2 \circ T_1 = 1$, but this is an unreasonable assumption.
For example, if $T_1$ decomposes a gate $G$, then $T_2$ must reassemble $G$ exactly.
Instead, we require that $T_2$ is a reflexive inverse to $T_1$.
The reflexive inverse is one of the many generalized inverses in algebra~\cite{BaksalaryTrenkler2021}, defined by the properties $T_1 \circ T_2 \circ T_1 = T_1$ and $T_2 \circ T_1 \circ T_2 = T_2$.
It follows immediately that $(T_2 \circ T_1)^2 = T_2 \circ T_1$ and $( T_1 \circ T_2 )^2 = T_1 \circ T_2$.
In other words, $T_2 \circ T_1$ and $T_1 \circ T_2$ are idempotent operators~\cite{MacLane2010}.
In practice, this means that translating back-and-forth many times between OpenQASM and Quipper is a stable operation, that will not induce arbitrarily large increases in file size.

We can refine these specifications by taking a closer look at each tool in LinguaQuanta.
We note that each tool is of the form $L_1 \xrightarrow{R} IR_1 \xrightarrow{T} IR_2 \xrightarrow{W} L_2$ where $R$ is a \emph{reader} that parses the input program, $T$ is a \emph{transformer} that rewrites the parsed input, and $W$ is a \emph{writer} that generates an output program from the result of $T$.
In the case where $(W \circ T \circ R)$ is \textsc{QasmToQuip}, we have that $L_1$ is the syntax for the OpenQASM language, $IR_1$ is an abstract representation of an OpenQASM program, $IR_2$ is an abstract representation of a Quipper program, and $L_2$ is the syntax for the Quipper ASCII format.

Returning to our original example, let $L_1 \xrightarrow{R_1} IR_1 \xrightarrow{T_1} IR_2 \xrightarrow{W_2} L_2$ denote \textsc{QasmToQuip} and $L_2 \xrightarrow{R_2} IR_2 \xrightarrow{T_2} IR_1 \xrightarrow{W_1} L_1$ denote \textsc{QuipToQASM}.
Then consider the equation $(W_1 \circ T_2 \circ R_2) \circ (W_2 \circ T_1 \circ R_1) = (W_1 \circ T_2 \circ T_1 \circ R_1)$.
This states that composing \textsc{QasmToQuip} with \textsc{QuipToQasm} is equivalent to composing the underlying transformers, as if the composition formed a single executable.
This is clearly a desirable property.
One way to ensure this property is to require that $R_1 \circ W_1 = 1$ and $R_2 \circ W_2 = 1$.
These new properties state that reading back what was previously written constitutes a no-op.
Categorically, $R_1$ is a retract of $W_1$ and $R_2$ is a retract of $W_2$~\cite{MacLane2010}

Then for any pair of LinguaQuanta tools $L_1 \xrightarrow{R_1} IR_1 \xrightarrow{T_1} IR_2 \xrightarrow{W_2} L_2$ and $L_2 \xrightarrow{R_2} IR_2 \xrightarrow{T_2} IR_1 \xrightarrow{W_1} L_1$, we require that the following rewrite rules hold.
\begin{enumerate}
    \item \textbf{Inversion (i)}. $T_i \circ T_j \circ T_i = T_i$ for $i, j \in \{ 1, 2 \}$ with $i \ne j$.
    \item \textbf{Retraction (i)}. $R_i \circ W_i = 1$ for $i \in \{ 1, 2 \}$.
\end{enumerate}
These specifications guide the design of each tool.

Of course, these specifications alone are not sufficient to ensure the correctness of LinguaQuanta.
For example, any pair of constant functions $C_1$ and $C_2$ satisfy the requirements of \textbf{Inversion (1)} and \textbf{(2)}.
Clearly, these specifications must be extended with a semantic notion of correctness.

For the purposes of a translations pipeline, it suffices to assume that both $L_1$ and $L_2$ have the same semantic domain.
Then there exists an object $SEM$ with morphisms $L_1 \xrightarrow{\sembrack{-}_1} SEM \xleftarrow{\sembrack{-}_2} L_2$ corresponding to the semantic interpretations of language $L_1$ and language $L_2$, respectively.
At the very least, each translation should preserve the semantic interpretation of its input.
In the case of \textsc{QasmToQuip}, this means that $\sembrack{-}_2 \circ W_2 \circ T_1 \circ R_1 = \sembrack{-}_1$.
Note, however, that these specifications do not ensure that reading an input, and then writing it back also preserves semantics.
For robust pipelines, this is also a reasonable assumption.
In the case of OpenQASM, this would mean that $\sembrack{-}_1 \circ W_1 \circ R_1 = \sembrack{-}_1$.
Based on these observations, we require that the following semantic rules hold.
\begin{enumerate}
    \setcounter{enumi}{2}
    \item \textbf{Preservation (i)}. $\sembrack{-}_j \circ W_j \circ T_i \circ R_i = \sembrack{-}_i$ for $i, j \in \{ 1, 2 \}$ with $i \ne j$.
    \item \textbf{Fluency (i)}. $\sembrack{-}_i \circ W_i \circ R_i = \sembrack{-}_i$ for $i \in \{ 1, 2 \}$.
\end{enumerate}
We refer to instances of rule $3$ as \textbf{Preservation} since they specify that each translation should preserve the semantic interpretation of its input text.
We refer to instances of rule $4$ as \textbf{Fluency} since they specify that each pair of readers and writers should maintain the semantic interpretation of their own language.
While \textbf{Fluency} is not \emph{necessary} for program correctness, it is likely the case that a lack of fluency indicates a yet undiscovered bug in the pipeline.

\section{Decompositions for Quantum Transpilation}
\label{Sect:Decompositions}

A large part of transpilation (i.e., defining each $T_i$) is to decompose gates from a source language into gates of a target language.
Naturally, this intersects with prior work on quantum circuit decomposition.
For simple gates, the required decompositions exist in the literature~\cite{AmyMaslov2013,Crooks2022,CrossJavadiAbhari2022,GilesSelinger2013,NielsenChuang2011,Selinger2013,Selinger2014}, as outlined in~\appendixcite{Appendix:Decomps}.
However, the general nature of control and rotation decomposition necessitates more general solutions.
In particular, we rely on the following four decomposition techniques to handle multi-controls and arbitrary rotations.

\begin{enumerate}
    \item Recall $e^{U^{\dagger}VU} = U^{\dagger}e^{V}U$ for any unitary $U$.
          Therefore, given any two rotations $R_A(\theta) = e^{iA\theta}$ and $R_B(\theta) = e^{iB\theta}$, if $B = V^{\dagger} A V$, then the following holds.
          \begin{center}
              \scalebox{0.85}{\begin{quantikz}
    & \gate{R_B(\theta)} &
\end{quantikz}}
=
\scalebox{0.85}{\begin{quantikz}
    & \gate{V} & \gate{R_A(\theta)} & \gate{V^{\dagger}} &
\end{quantikz}}

          \end{center}
          This provides a change of basis between rotation operators.
    \item Assume that $U = CXBXA$ and $CBA = I$.
          Then the following holds by~\cite{BarencoBennett1995}.
          \begin{center}
              \scalebox{0.85}{\begin{quantikz}
    & \ctrl{1} & \\
    & \gate{U} &
\end{quantikz}}
=
\scalebox{0.85}{\begin{quantikz}
    & & \ctrl{1} & & \ctrl{1} & & \\
    & \gate{A} & \targ{} &  \gate{B} & \targ{} & \gate{C} &
\end{quantikz}}
          \end{center}
          Let $R(\theta)$ be a rotation and $D$ self-inverse.
If $V = DXD$,  $V \cdot R(\theta) \cdot V = R(-\theta)$, $R(0) = I$, and $R(\theta) \cdot R(\gamma) = R(\theta + \gamma)$, then the following holds by taking $U = R(\theta)$, $C = D$, $B = D \cdot R(-\theta) \cdot D$, and $A = D \cdot R(\theta)$.
          \begin{center}
              \scalebox{0.85}{\begin{quantikz}
    & \ctrl{1} & \\
    & \gate{R(\theta)} &
\end{quantikz}}
=
\scalebox{0.85}{\begin{quantikz}
    & & & \ctrl{1} & & & & \ctrl{1} & & \\
    & \gate{R \left( \frac{\theta}{2} \right)} & \gate{D} & \targ{} & \gate{D} & \gate{R \left( -\frac{\theta}{2} \right)} & \gate{D} & \targ{} & \gate{D} &
\end{quantikz}}
          \end{center}
          This provides decompositions for many rotations with single controls.
    \item A \emph{Toffoli-like gate} is a three-qubit unitary $U$ equipped with two-qubit states $\ket{\varphi_1}$ through to $\ket{\varphi_4}$ such that the following equations hold.
          \begin{align*}
              U \ket{000} &= \ket{\varphi_1}\ket{0}
              & U \ket{100} &= \ket{\varphi_2}\ket{0}
              \\
              U \ket{010} &= \ket{\varphi_3}\ket{0}
              & U \ket{110} &= \ket{\varphi_4}\ket{1}
          \end{align*}
          If $U$ is Toffoli-like, then the following holds by~\cite{Selinger2013}.
          \begin{center}
              \scalebox{0.85}{\begin{quantikz}
    & \ctrl{1} & \\
    & \ctrl{1} & \\
    & \gate{G} & \\
\end{quantikz}}
=
\scalebox{0.85}{\begin{quantikz}[wire types={q,q,n,q}]
    & & \gate[3]{\;U\;} & & \gate[3]{U^\dagger} & & \\
    & & & & & & \\
    & \lstick{\ket{0}} & \setwiretype{q} & \ctrl{1} & & \rstick{\ket{0}} & \setwiretype{n} \\
    & & & \gate{G} & & &
\end{quantikz}}
          \end{center}
          This decomposition can be applied until only a single control remains.
          Note that the choice of $U$ is not arbitrary, and may impact circuit efficiency.
    \item Assume that $U = V^2$.
          Then the following holds by~\cite{BarencoBennett1995,SleatorWeinfurter1995}.
          \begin{center}
              \scalebox{0.85}{\begin{quantikz}
    & \ctrl{1} & \\
    \ghost{X} & \ctrl{1} & \\
    & \gate{U} &
\end{quantikz}}
=
\scalebox{0.85}{\begin{quantikz}
    & & \ctrl{1} & & \ctrl{1} & \ctrl{2} & \\
    \ghost{X} & \ctrl{1} & \targ{} & \ctrl{1} & \targ{} & & \\
    & \gate{V} & & \gate{V^{\dagger}} & & \gate{V} &
\end{quantikz}}
          \end{center}
\end{enumerate}
Note that Decomposition $2$ also applies to user-defined rotations, provided that the end-user specifies $D$.
In the current version of LinguaQuanta, the list of supported rotations, and their corresponding $D$ operators, is hard-coded.

\section{Challenges Faced and Proposed Solutions}
\label{Sect:Challenges}

Many aspects of LinguaQuanta are routine, such as rewriting gates via known decompositions, or inlining built-in OpenQASM 3 functions.
However, certain differences between OpenQASM 3 and Quipper give rise to design problems for which the solutions are far less obvious.
In this section, we state these design challenges, and summarize our proposed solutions.

\paragraph{Approximating the Identity on Round Translation.}
As outlined in \cref{Sect:Pipeline}, it is unreasonable to require that LinguaQuanta acts as the identity function on round translations.
However, it is still desirable to minimize the distance between the round translation function and the identity function.
One reason why round translations fail to act as the identity function is that the decomposition of unitary operators is not invertible.
For this reason, we attempt to minimize the number of unitary decompositions in LinguaQuanta.
In particular, we have implemented every Quipper gate in an OpenQASM library named \lstinline{quipgates}.
Furthermore, we have backported all OpenQASM 3 gates to OpenQASM 2.0, so that neither Quipper gates nor OpenQASM 3 gates require decompositions when translated to OpenQASM 2.0.
By minimizing the number of decompositions in LinguaQuanta, we also reduce the risk of faults in our implementation (i.e., smaller code surface), while also improving traceability when debugging LinguaQuanta.
These techniques should be applicable to any pair of quantum programming languages with distinct gate sets, and support for user-defined unitary operators.

\begin{figure}[t]
    \centering
    \includegraphics[scale=0.8]{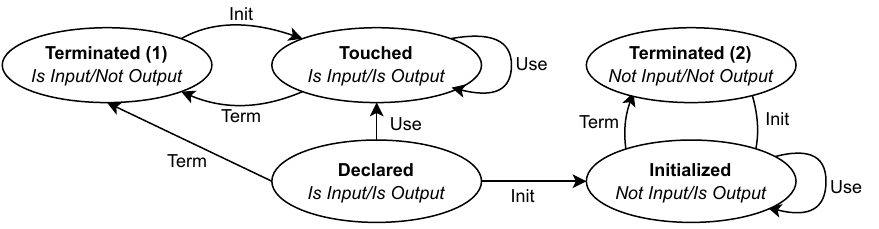}
    \caption{A DFA to infer the state of a Quipper wire.
             The alphabet consists of Init, Term, and Use.
             Missing edges indicate one-way transitions to an error state.}
    \label{Fig:WireDFA}
\end{figure}

\paragraph{Translating OpenQASM Measurements to Quipper.}
Recall from \cref{Sect:Background:Lang}, that OpenQASM 3 and Quipper provide different primitives for qubit measurement.
In OpenQASM, the measurement operator produces a classical bit as output, and collapses the state of a quantum register as a side-effect.
In contrast, the measurement operator in Quipper remains side-effect free by first collapsing the state of a quantum wire, and then casting the collapsed quantum wire to a classical wire.
It is clear that the semantics of OpenQASM subsumes the semantics of Quipper.
To capture the semantics of OpenQASM in Quipper, we introduce the following circuit.
\begin{center}
    \begin{quantikz}[wire types={q,n}]
    & & \ctrl{1} & & \\
    &\lstick{\ket{0}} & \targ{}\setwiretype{q} & \meter{} &\setwiretype{c}
\end{quantikz}
\end{center}
This circuit first prepares an entangled pair of qubits from the wire under measurement, and then measures the second wire in the pair to produce a classical wire as output while collapsing the first wire via entanglement.

\paragraph{Typed Wires and Ancillas in OpenQASM.}
Recall from \cref{Sect:Background:Lang}, that the Quipper circuit model allows for new wires to be allocated (or deallocated) on-the-fly via ancilla management.
However, OpenQASM does not support ancilla qubits.
Therefore, we have implemented an OpenQASM 3 function with the same semantics as the ancilla allocation and deallocation gates in Quipper.
Our goal is to recover all ancilla management data, including the total number of wires in use at the start and end of the circuit.
In addition to better approximating a round translation, this also allows us to validate the ancilla management code generated by LinguaQuanta, which facilitates bug detection in LinguaQuanta.
To this end, we associate with each register in OpenQASM a deterministic finite-state machine (DFA) of the form shown in \cref{Fig:WireDFA}.
This DFA tracks the allocation state of each conceptual wire, and detects invalid ancilla management (e.g., double initialization, or use before initialization).
Essentially, this collection of DFA's form a product DFA indexed by qubits, where the state of each component depends only on the operations that act locally on the associated qubit.
In addition to ancilla management, Quipper also supports type-conversion, which is unsupported in OpenQASM.
For example, measurement in Quipper casts a quantum wire to a classical wire, whereas measurement in OpenQASM simple produces a new classical register.
To overcome this limitation, we make use of shadow memory, as commonly seen in classical program instrumentation~\cite{SewardNethercote2005}.
At a high-level, shadow memory allows more than one memory cell to be associated with a single memory address, often to store metadata.
In the case of LinguaQuanta, every quantum register can be shadowed by a classical register.
Dually, every classical register can be shadowed by a quantum register.
These registers are allocated on-the-fly to minimize memory overhead, but are never reclaimed by LinguaQuanta.

\paragraph{Ensuring Controls Remain Inlined.}
Since OpenQASM 3 and Quipper both support controlled unitary operators, then translating controls between these two languages is straight-forward.
However, the OpenQASM 2.0 language does not offer such a feature.
This means that when translating from OpenQASM 3 to OpenQASM 2.0, it is necessary to first eliminate all control modifiers.
However, the Quipper library already implements most of the control decompositions required for OpenQASM 2.0 legacy support.
To avoid rewriting well-tested code, LinguaQuanta performs all control elimination on the level of Quipper programs.
This means that control elimination is not possible on the side of OpenQASM.
If a new control is introduced during or after the translation from Quipper to OpenQASM, there is no way to eliminate the control.
This means that in addition to our categorical requirements, we must also impose the syntactic requirement that pipeline components do not introduce new controls.
Here, a control is called \emph{new}, if an uncontrolled gate is translated to one or more controlled gates.
Currently, this requirement is enforced manually.
In the future, we would like to provide some guarantee for this property, either through formal verification or light-weight formal methods.

\section{LinguaQuanta by Example}
\label{Sect:Tutorial}

The design of LinguaQuanta was motivated by real-world examples in quantum program compilation and analysis.
In this section, we begin by discussing once such real-world example.
Through this example, each tool in the LinguaQuanta pipeline is motivated and described.
Following this example, the central tools \textsc{QasmToQuip} and \textsc{QuipToQasm} are explored in more detail, using the QPE algorithm introduced in \cref{Sect:Background:Lang}.
Further examples and explanation can be found in the LinguaQuanta documentation.

\subsection{An Example Pipeline in LinguaQuanta}

In recent work by LeBlonde et. al~\cite{Leblond2023}, a prototype of LinguaQuanta was used to compile Quipper programs to fault-tolerant quantum hardware.
The goal of this paper was to estimate the cost of running certain quantum chemistry simulations on realistic quantum hardware.
The simulations were written in Quipper using the NewGSE techniques described in~\cite{KornellSelinger2023}.
However, the fault-tolerant hardware targeted by LeBlonde et. al makes use of surface codes, for which operations are performed using lattice surgery instructions rather than high-level quantum gates~(see~\cite{WatkinsNguyen2023}, for an introduction to logical lattice instructions).
Typically, the Lattice Surgery Compiler (LSC)~\cite{WatkinsNguyen2023} is used to compile to this lattice surgery instruction set.
However, the LSC only supports a subset of OpenQASM 2.0 as its input language.
To bridge this gap, LeBlonde et. al made use of LinguaQuanta to translate each simulation to the LSC subset of the OpenQASM 2.0 language.

\begin{figure}[t]
    \centering
    \includegraphics[scale=0.7]{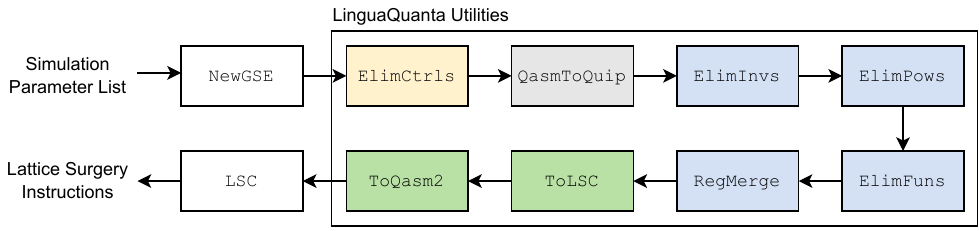}
    \caption{The translation pipeline implemented in~\cite{Leblond2023}. In this pipeline orange boxes indicate pre-processing, blue boxes indicate post-processing, and green boxes indicate additional steps for legacy support.}
    \label{Fig:Pipeline}
\end{figure}

The pipeline used by LeBlonde et. al can be found in~\cref{Fig:Pipeline}.
Recall from \cref{Sect:Challenges} that OpenQASM 2.0 does not support control modifiers.
Similarly, OpenQASM 2.0 does not support inverse modifiers, power modifiers, or function calls.
As outlined in \cref{Sect:Challenges}, control elimination is performed on the level of Quipper programs, since the Quipper library implements many of the control decompositions necessary for OpenQASM 2.0 support.
For this reason, the pre-processing stage in the pipeline applies \textsc{ElimCtrls} to obtain a Quipper program with only the supported controlled operations.
Next, \textsc{QasmToQuip} is applied to obtain an OpenQASM 3 program without any controls.
The post-processing on this OpenQASM 3 program begins by applying \textsc{ElimInvs}, \textsc{ElimPows}, and \textsc{ElimFuns} to obtain a valid OpenQASM 2.0 program.
Next, \textsc{RegMerge} is applied to obtain a valid LSC program\footnote{The LSC requires that all qubits (resp.~cbits) appear as a single array.}
Following the post-processing stage, \textsc{ToLSC} and \textsc{ToQasm2} are applied to reformat the output syntax to conform with the LSC subset of the OpenQASM 2.0 language.
The full pipeline is expanded in more detail by the \texttt{quip\_to\_lsc.sh} script in LinguaQuanta.

The key step in~\cref{Fig:Pipeline} is to translate the pre-processed Quipper into a valid OpenQASM 3 program.
In this sense, \textsc{QuipToQasm} is the most important LinguaQuanta tool featured in~\cref{Fig:Pipeline}.
More generally, all LinguaQuanta pipelines will rely on either \textsc{QuipToQasm} or \textsc{QasmToQuip} to realize a translation between the two languages.
For this reason, it is worthwhile exploring \textsc{QuipToQasm} and \textsc{QasmToQuip} in more detail.
As outlined in \cref{Sect:Pipeline}, these tools are reflexive inverses which compose to an idempotent operator.
To illustrate these tools and their properties, the concrete example of QPE is explored.

\begin{landscape}
    \begin{figure}[t]
    \centering
    \setlength{\columnsep}{-2.5cm}
    \parbox{\textwidth}{
\begin{lstlisting}[style=qasm,
                     multicols=2,
                     basicstyle=\tiny\ttfamily]
OPENQASM 3;
include "stdgates.inc";
include "quipfuncs.inc";

qubit[1] input_qwires;
qubit qtmp_0; QInit0(qtmp_0);
qubit qtmp_1; QInit0(qtmp_1);
qubit qtmp_2; QInit0(qtmp_2);

h qtmp_0; h qtmp_1; h qtmp_2;

ctrl @ t qtmp_0, input_qwires[0];
ctrl @ t qtmp_0, input_qwires[0];
ctrl @ t qtmp_0, input_qwires[0];
ctrl @ t qtmp_0, input_qwires[0];
ctrl @ t qtmp_1, input_qwires[0];
ctrl @ t qtmp_1, input_qwires[0];
ctrl @ t qtmp_2, input_qwires[0];

h qtmp_0; cp(¤$\pi$¤/4.0) qtmp_1, qtmp_0; cp(¤$\pi$¤/8.0) qtmp_2, qtmp_0;
h qtmp_1; cp(¤$\pi$¤/4.0) qtmp_2, qtmp_1;
h qtmp_2;

bit ctmp_3; ctmp_3 = QMeas(qtmp_0);
bit ctmp_4; ctmp_4 = QMeas(qtmp_1);
bit ctmp_5; ctmp_5 = QMeas(qtmp_2);

ctmp_3 = CDiscard(); ctmp_4 = CDiscard(); ctmp_5 = CDiscard();
\end{lstlisting}}
    \caption{A translation of \cref{Fig:QuipQPE:Orig} using \textsc{QuipToQasm}.
             Variables have been renamed.}
    \label{Fig:QuipQPE:Qasm}
    \setlength{\columnsep}{0cm}
\end{figure}

\begin{figure}[t]
    \centering
    \includegraphics[scale=1.2]{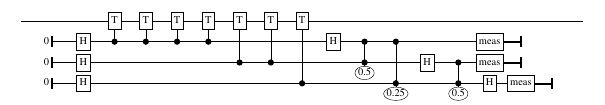}
    \caption{A round-trip translation of \cref{Fig:QuipQPE:Orig}.}
    \label{Fig:QuipQPE:Round}
\end{figure}

\begin{figure}[t]
    \centering
    \includegraphics[scale=1.2]{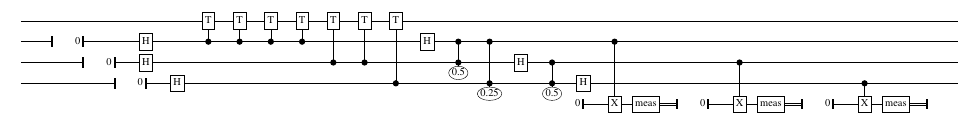}
    \caption{A translation of \cref{Fig:QasmQPE:Orig} using \textsc{QasmToQuip}.
             Note that in Quipper $X$ is used rather than $\oplus$.}
    \label{Fig:QasmQPE:Quip}
\end{figure}
\end{landscape}

\subsection{Round Translation of the QPE Algorithm}

Let $U$ be a unitary operator with eigenvector $\ket{\psi}$ and associated eigenvector algorithm.
Recall from \cref{Sect:Background:Lang}, that the QPE algorithm~\cite{Kitaev1995} applies the $k$-qubit quantum Fourier transform~\cite{Coppersmith1994} to determine $\theta$ from $U$ and $\ket{\psi}$ with an accuracy of $k$ binary digits.

As in \cref{Fig:QasmQPE:Orig} and \cref{Fig:QuipQPE:Orig}, we will consider the case where $U$ is the unitary operator $T$ with eigenvector $\ket{1}$ and associated eigenvalue $e^{i\pi/4}$.
Then given $k = 3$, the QPE algorithm will produce an estimate of $\pi/4$ up to $3$ binary digits.
The algorithm can be broken down into three stages.
In the first stage (see lines~$6$--$8$ in \cref{Fig:QasmQPE:Orig}), $3$ qubits are prepared in a uniform superposition (i.e.,~a state from which all $3$-qubit sequences can be observed with equal probability).
In the second stage (see lines~$10$--$12$ in \cref{Fig:QasmQPE:Orig}), the $T$ operator is controlled by each of these $3$ qubits, and applied to $\ket{\phi}$ exactly $2^{2-j}$ times, where $j \in \{ 0, 1, 2 \}$ is the index of the qubit.
In the third stage (see lines~$14$--$16$ in \cref{Fig:QasmQPE:Orig}), the quantum Fourier transform is applied to the $3$ qubits, to obtain a state whose measurement outcome corresponds to an estimate of $\pi/4$ up to $3$ binary digits.
These three stages can be clearly visualized in \cref{Fig:QuipQPE:Orig}.

Starting from \cref{Fig:QuipQPE:Orig}, an OpenQASM 3 program is obtained by running \textsc{QuipToQasm}.
The output is illustrated in \cref{Fig:QuipQPE:Qasm}.
Superficially, this output is very similar to the handcrafted OpenQASM program illustrated in \cref{Fig:QasmQPE:Orig}.
\emph{In particular, the three phases of the QPE algorithm are still distinguishable.}
However, there are two major differences.
First, note the new function calls \lstinline[style=qasm]{QInit0}, \lstinline[style=qasm]{QMeas}, and  \lstinline[style=qasm]{CDiscard}.
These functions are defined in LinguaQuanta's \lstinline{quipfuncs.inc} library, and allow LinguaQuanta to encode ancilla management metadata within an OpenQASM 3 program.
Next, note the \lstinline[style=qasm]{input_qwires} register.
This is used to encode all input wires from the Quipper circuit.
Then, for each ancilla, a variable prefixed by \lstinline{qtmp} is introduced.
These registers stay associated with the wires throughout the entire translation.
The first time a type conversion occurs (e.g., at each measurement), a new register is declared, which becomes associated with the wire.
In this example, all such registers are prefixed by \lstinline{ctmp}.
The pair of \lstinline{qtmp} and \lstinline{ctmp} form the shadow memory outlined in \cref{Sect:Challenges}.
Note that LinguaQuanta does not depend on variable names to recover ancilla data.

To illustrate a round translation, \textsc{QasmToQuip} is then applied to the program in \cref{Fig:QuipQPE:Round}.
It is interesting to note that the circuit in \cref{Fig:QuipQPE:Round} is almost identical to the circuit in \cref{Fig:QuipQPE:Orig}.
Ancilla recovery in \cref{Fig:QuipQPE:Round} was facilitated by the ancilla management functions in \cref{Fig:QuipQPE:Qasm}, together with the ancilla DFA outlined in \cref{Fig:WireDFA}.
The one difference to note is that the $R(2\pi/k)$ gates in \cref{Fig:QasmQPE:Orig} have been replaced by doubly controlled phase gates in \cref{Fig:QuipQPE:Round}.
This is because $R(2\pi/k)$ is equivalent to a controlled global phase of $2\pi/k$.

Of course, it is also possible to translate an OpenQASM 3 program to a Quipper program.
However, this direction often results in larger changes to the source text.
For example, the Quipper program in \cref{Fig:QasmQPE:Quip} is obtained by applying \textsc{QasmToQuip} to \cref{Fig:QasmQPE:Orig}.
Qualitatively, there are two large changes to note.
First, each \lstinline[style=qasm]{reset} in the OpenQASM 3 program has been replaced by qubit termination, followed immediately by qubit initialization.
This is equivalent to resetting a qubit.
Second, all measurements have been replaced by the measurement circuits presented in \cref{Sect:Challenges}.
Note that the results are immediately discarded, since the measurement results are never used by the original OpenQASM program.

\section{Discussion and Related Work}
\label{Sect:Discussion}

To the best of our knowledge, we are the first to develop a quantum transpiler with the goal of interoperability, rather than circuit optimization.
However, LinguaQuanta is not the first tool to translate between OpenQASM and Quipper.
For example, the PyZX circuit optimizer~\cite{KissingerWetering2020} is capable of taking OpenQASM and Quipper programs, both as inputs and as outputs.
However, the language support in PyZX is minimal, with many gates unsupported, and no plans for measurement and classical control.
Furthermore, tools such as PyZX rely on circuit extraction, and therefore make no guarantee about preserving the structure of the input program.
In this respect, LinguaQuanta is the first tool of its kind.

A major challenge faced in the development of LinguaQuanta was support for parameterized unitary gates, otherwise known as \emph{rotation gates} in Quipper.
In both Quipper and OpenQASM, users may define their own so-called rotation gates.
In OpenQASM, this is achieved through the composition of single-qubit rotations and controlled-not gates.
In Quipper, rotations are simply one-parameter opaque gates.
It is not hard to see that these rotation gates may not define a rotation about a \emph{fixed axis}.
This freedom often contradicts the assumptions a developer might make about rotation gates.
For example, one might expect that given a rotation gate $R(\theta)$, the relation $R(\theta) \cdot R(\gamma) = R(\theta + \gamma)$ always holds.
However, the $R$ gate in Quipper, as taken from the quantum Fourier transform~\cite{GreenLumsdaine2013b}, instead satisfies $R(\theta^{-1}) \cdot R(\gamma^{-1}) = R((\theta + \gamma)^{-1})$.
From a compilation and program analysis point-of-view, this lack of standardization is a serious concern.

We note that an important development in classical program design was the introduction of abstract data types (ADTs)~\cite{LiskovZilles1974}.
ADTs allow developers to work with common data types, such as lists and arrays, without concern for implementation details.
In turn, these ADTs can then be used by compilers and tools to reason compositionally about the program under analysis.
For example, consider the gate $C( R( \theta ) )$ where $R( \theta )$ is an unknown unitary parameterized by an angle $\theta \in [0, 2 \pi)$.
One might hope that Decomposition~2 in \cref{Sect:Decompositions} would yield an implementation for $C( R( \theta ) )$ using the standard gate set and $R( \theta )$.
However, this decomposition requires that $R(0) = I$, $R(\theta) \cdot R(\gamma) = R(\theta + \gamma)$, and $V \cdot R(\theta) \cdot V = R(-\theta)$ for some self-inverse unitary $V$.
Therefore, a compiler would require further specifications about $R(\theta)$ for the decomposition to be possible.
Based on such examples, we argue that quantum programming languages are in need of \emph{abstract rotation types}.
Abstract rotation types would provide a taxonomy for the parameterized unitary gates used in quantum programming, together with first-order specifications.
These types would act as a contract between the developer and the compiler, allowing for better optimizations.
Based on our experiments, we start this discussion by proposing a list of abstract specifications.
\begin{enumerate}
    \item \textbf{Zero is Identity}:
          $R(0) = I$.
    \item \textbf{Negation is Inversion}:
          $R(\theta) \cdot R(-\theta) = I$, or $R(\theta)^{-1} = R(-\theta)$.
    \item \textbf{Negation by $G$-Conjugation}:
          $G^{-1} \cdot R(\theta) \cdot G = R(-\theta)$ for some gate $G$.
    \item \textbf{Commutativity}:
          $R(\theta) \cdot R(\gamma) = R(\gamma) \cdot R(\theta)$.
    \item \textbf{Additivity}:
          $R(\theta) \cdot R(\gamma) = R(\theta + \gamma)$.
    \item \textbf{Inverse Addivitity}:
          $R(\theta^{-1}) \cdot R(\gamma^{-1}) = R((\theta + \gamma)^{-1})$.
    \item \textbf{$p$-Periodicity}:
          $R(\theta) = R(\theta + kp)$ for all $k \in \mathbb{N}$.
\end{enumerate}
Dependencies between these properties suggest a hierarchy of abstract rotations.
This hierarchy differs from equation theories of quantum circuits (e.g.,~\cite{Clement2023}), since abstract rotations may abstract away details such as the axes of rotation.

Whereas the lack of abstract rotation types is a limitation of all languages, several other limitations we encountered were entirely language-specific.
For example, the Quipper framework can generate self-inverse gates with inverse modifiers, but will eliminate those same modifiers when reading back the circuit.
This means that the reader is not a retraction of the writer in Quipper.
From a semantic perspective, this is inconsequential, but may lead to unexpected behaviour when fit into a program analysis pipeline.
Furthermore, whereas Quipper-related tools support angles of arbitrary precision~\cite{RossSelinger2016}, the internal representation used by Quipper is double-precision.
For many applications, double-precision is insufficient, though an end-user be unaware of this discrepancy.
Finally, we note the lack of ancillas in OpenQASM.
The reason stated for this design choice is that qubit allocation is too expensive at runtime.
However, OpenQASM is intended as a language for all stages of quantum compilation, including algorithm specification.
Thus, OpenQASM couples ancilla management with algorithm specification, and fails to separate concerns.
As shown in this paper, recursion-free ancilla allocation can be resolved at compile-time.
An interesting direction for future work is to reason about ancillas in the presence of recursion at compile-time.

The current version of LinguaQuanta is limited to the features required to compile Quipper programs to surface code instructions via the OpenQASM tool in~\cite{WatkinsNguyen2023}.
In future work, we plan to extend and improve the design and implementation of LinguaQuanta as follows:
\begin{enumerate}
\item Support for hierarchically-defined and opaque unitary gates;
\item Improved power modifier support;
\item Constant value propagation and bounded iteration;
\item White-box pipeline fuzzing and empirical evaluations;
\item Refined categorical foundations.
\end{enumerate}
See \appendixcite{Appendix:Future} for further details.

%

\section{Acknowledgements}
\label{sec:ack}

We would like to thank {Xiaoning Bian} for his help during the early development stages of LinguaQuanta, and {Peter Selinger} for his perspectives on the categorical specification of LinguaQuanta.
This research was sponsored in part by \redact{the United States Defense Advanced Research Projects Agency (DARPA) under the Quantum Benchmarking program, contract \# HR001122C0066}.

\bibliographystyle{plain}
\bibliography{bibliography}

\addappendix{\newpage
\appendix
\allowdisplaybreaks

\section{Matrix Exponentiation, Rotations, and Global Phases}
\label{Appendix:RotBG}

We assume that the reader is familiar with basis complex analysis.
It is a well-known result that the Maclaurin series for $e^z$ is $\sum_{n=1}^{\infty} \frac{1}{n!} z^n$ with the series converging to $e^z$ for all $z \in \mathbb{C}$~(see,~e.g.,~\cite{Hall2000}).
From this definition, we obtain an extension of exponentiation to complex matrices.
If $M$ is a square complex matrix, then we define $e^M$ to be $\sum_{n=1}^{\infty} \frac{1}{n!} M^n$.
The matrix exponential has many counter-intuitive properties~(see,~e.g.~\cite{Hall2000}).
For example, if $M$ and $N$ are both square matrices of the same size, then $e^{M+N} = e^N e^M$ if and only if $MN = NM$.
Despite these surprising properties, matrix exponentiation proves to be an invaluable tool in the theory of quantum computation.
For example, if $G \in \mathcal{U}(2)$ and $G^\dagger = G$, then $e^{iG\theta}$ can be interpreted as a rotation by $\theta$ about some axis defined by $G$.
\section{Outline for Future Work}
\label{Appendix:Future}

\paragraph{Improved Power Modifier Support.}
Many unitary operators generate groups of finite order.
For example, self-inverse gates have order $2$ and the $iX$ gate has order $4$.
Furthermore, all periodic rotations with angles taken to be rational multiples of their periods must also be of finite order.
If a unitary $G$ has order $k$, then power modifiers may be applied to $G$ modulo $k$.
This optimization would reduce blow-up in program size from large power modifiers.

\paragraph{Hierarchically Defined Unitary Gates.}
OpenQASM 3 supports hierarchically defined unitary gates.
These are named unitary gates, defined in terms of other unitary gates.
To translate such a gate from OpenQASM 3 to Quipper, it suffices to inline each definition.
To inline the inverse of a hierarchically defined unitary gate, one must simply note that given unitary gates $U$ and $V$, $(UV)^{\dagger} = V^{\dagger} U^{\dagger}$.
In both cases, inlining would increase the distance between round translations and the identity function.
However, this limitation is unavoidable, since hierarchically defined unitary gates are not supported in Quipper.
One interesting question is whether to inline immediately or not.
If the gates are not inlined immediately, then their compositional definitions simplify higher-level reasoning, such as canceling inverse gates or reasoning about gate orders.

\paragraph{Opaque Unitary Gates.}
Both Quipper and OpenQASM support opaque gates.
An opaque gate is a unitary gate with an unknown implementation.
This is analogous to external functions in classical programming.
To support opaque gates, there are two challenges to overcome.
Firstly, opaque gates are declared by first-use in Quipper, and must be inferred from the source text.
Secondly, the syntax for opaque gates differs significantly between OpenQASM 2.0 and 3.
Neither challenge is fundamentally hard, but does require engineering effort.

\paragraph{Constant Value Propagation.}
LinguaQuanta currently requires that runtime constants in OpenQASM 3 are in fact literal expressions.
This is a very restrictive assumption, and in fact encourages bad programming practices such as magic numbers~\cite{LimaSouza2020}.
To support programs with constant variable, we plan to implement basic data-flow analysis to achieve constant propagation~\cite{KamUllman1997}.

\paragraph{Handling Bounded Iteration.}
OpenQASM 3 supports loops and bounded iteration over quantum registers.
In the future, we plan for LinguaQuanta to support bounded loops and iteration over quantum registers.
This is a well-known problem, and can be solved via loop unrolling~\cite{Aho1977}. 

\paragraph{White-Box Pipeline Fuzzing.}
Categorical specifications open many new directions for dynamic program analysis.
A standard technique in dynamic program analysis is random testing, otherwise known as fuzzing~\cite{MillerFredriksen1990}.
Prior work has used white-box fuzzing to explore the input syntax of compilers to improve test coverage with little manual effort~\cite{GodefroidKiezun2008}.
These techniques should also apply to the languages defined by pipeline rewrite rules.
Then standard fuzzing techniques could search for pairs of distinct pipelines, for which the rewrite rules assert the pipelines must be equivalent, yet concrete inputs demonstrate otherwise.
This would indicate that the transpiler violates one of its categorical specification.

\paragraph{Refined Categorical Foundations.}
This paper has laid out a groundwork for categorically specified software pipelines.
However, some subtle points were not addressed, such as how idempotents compose with retracts, and what this composition tells us about software pipeline design.
Furthermore, the rewrite rules in \cref{Sect:Pipeline} assume that all pipelines are composed from only two tools.
The rewrite rules generalize easily if we fix the order of composition (for example, via typing).
However, several tools in the LinguaQuanta pipeline are in fact endomorphisms, in that their input and output language are the same.
Building a better understanding of these endomorphic tools is an important question for future work.
In terms of semantic correctness, it remains an open problem to formally specify both $\sembrack{-}_1$ and $\sembrack{-}_2$ for LinguaQuanta, and to verify both \textbf{Preservation} and \textbf{Fluency} formally.
Finally, it would be interesting to study the failure for round translations to compose to the identity function.
For example, given an adequate choice of posetal categories, it might be possible to study this failure through Galois connections and closure operators.

\paragraph{Empirical Evaluations.}
In our limited experience, LinguaQuanta scales well to large-scale quantum programs.
Minor performance issues have been identified in the Quipper ASCII parser, though this is unsurprising, since the Quipper ASCII parser was designed for small input programs.
To better understand the impact of these performance issues, and whether it would be worthwhile improving the Quipper ASCII parser, we plan to evaluate the performance of LinguaQuanta on a collection of large-scale quantum programs.
 We would also like to study the number of real-world OpenQASM and Quipper programs that fall within the feature set supported by LinguaQuanta.

\paragraph{Out-Of-Scope Features.}
We do not support dynamic lifting~\cite{GreenLumsdaine2013a} or no-control flags in Quipper, as we do not target dynamic circuit generation.
We do not support low-level features in OpenQASM 3 such as physical qubits and the OpenPulse language~\cite{CrossJavadiAbhari2022}, nor do we support very high-level features such as runtime floating-point arithmetic and dynamic arrays, since the current scope of LinguaQuanta is logical circuit translation.
We currently have no plans to support generalized controls in Quipper, since generalized controls are simply hints to a program analyzer, and do not have an analogue in OpenQASM. 

\section{Decomposing U-like Gates in OpenQASM}
\label{Sect:UGate}
\label{Appendix:U}

All unitary gates in OpenQASM 3 are defined using a three parameter unitary gate $U(\theta, \phi, \lambda)$~\cite{CrossJavadiAbhari2022}.
The parameters $\theta$, $\phi$, and $\lambda$ all range from $[0, 2\pi)$.
The definition of $U(\theta, \phi, \lambda)$ is as follows.
\begin{equation} \label{Eqn:Universal}
    U(\theta, \phi, \lambda) = 
    \begin{bmatrix}
      \cos(\theta / 2)            & -e^{i\lambda} \sin(\theta / 2) \\
      -e^{i\phi} \sin(\theta / 2) & e^{i(\phi+\lambda)} \cos(\theta / 2)
    \end{bmatrix}
\end{equation}
Define $P(\alpha) = \diag( 1, e^{i\alpha} )$.
In \cite{CrossJavadiAbhari2022}, this gate is used to provide a useful decomposition of \cref{Eqn:Universal}, as depicted in \cref{Eqn:DecompPH}.
\begin{equation} \label{Eqn:DecompPH}
    U(\theta, \phi, \lambda) = e^{i\theta/2} \cdot P(\phi + \pi / 2) \cdot H \cdot P(\theta) \cdot H \cdot P(\lambda - \pi / 2)
\end{equation}
Furthermore,
\begin{equation} \label{Eqn:PRz}
    P(\alpha) = \diag\left(1, e^{i\alpha} \right) = e^{i\alpha/2} \cdot \diag\left( e^{-i\alpha/2}, e^{i\alpha/2} \right) = e^{i\alpha/2} \cdot e^{i(-\alpha/2)Z}
\end{equation}
Note that $P(\pi/2) = \diag( 1, i ) = S$ and $P(-\pi/2) = \diag( 1, -i ) = S^{\dagger}$.

\subsection{Decomposing U and U3 Gates}

In OpenQASM 3, the \lstinline[style=qasm]{U3($\theta$,$\phi$,$\lambda$)} gate is defined as \lstinline[style=qasm]{gphase(-($\phi$+$\lambda$)/2)} followed by \lstinline[style=qasm]{U($\theta$,$\phi$,$\lambda$)}.
Then it suffices to give a translation for \lstinline[style=qasm]{U($\theta$,$\phi$,$\lambda$)}.
Then by \cref{Eqn:DecompPH,Eqn:PRz}, \lstinline[style=qasm]{U($\theta$,$\phi$,$\lambda$)} decomposes as follows.
\begin{align*}
    U(\theta, \phi, \lambda) &= e^{i\theta/2} \cdot P(\phi + \pi / 2) \cdot H \cdot P(\theta) \cdot H \cdot P(\lambda - \pi / 2) \\
    &= e^{i\theta/2} \cdot P(\phi) \cdot P(\pi / 2) \cdot H \cdot P(\theta) \cdot H \cdot P(-\pi/2) \cdot P(\lambda) \\
    &= e^{i\theta/2} \cdot P(\phi) \cdot S \cdot H \cdot P(\theta) \cdot H \cdot S^{\dagger} \cdot P(\lambda) \\
    &= e^{i\theta/2} \cdot P(\phi) \cdot S \cdot H \cdot \left( e^{i\theta/2} \cdot R_z(\theta) \right) \cdot H \cdot S^{\dagger} \cdot P(\lambda) \\
    &= e^{i\theta} \cdot P(\phi) \cdot S \cdot H \cdot R_z(\theta) \cdot H \cdot S^{\dagger} \cdot P(\lambda) \\
    &= e^{i\theta} \cdot \left( e^{i\phi/2} \cdot R_z(\phi) \right) \cdot S \cdot H \cdot R_z(\theta) \cdot H \cdot S^{\dagger} \cdot P(\lambda) \\
    &= e^{i(\theta+\phi/2)} \cdot R_z(\phi) \cdot S \cdot H \cdot R_z(\theta) \cdot H \cdot S^{\dagger} \cdot P(\lambda) \\
    &= e^{i(\theta+\phi/2)} \cdot R_z(\phi) \cdot S \cdot H \cdot R_z(\theta) \cdot H \cdot S^{\dagger} \cdot \left( e^{i\lambda/2} \cdot R_z(\lambda) \right) \\
    &= e^{i(\theta+\phi/2+\lambda/2)} \cdot R_z(\phi) \cdot S \cdot H \cdot R_z(\theta) \cdot H \cdot S^{\dagger} \cdot R_z(\lambda)
\end{align*}
In conclusion:
\begin{equation} \label{Eqn:UDecomp}
    U(\theta, \phi, \lambda) = e^{i\theta} \cdot e^{i(\phi+\lambda)/2} \cdot R_z(\phi) \cdot S \cdot H \cdot R_z(\theta) \cdot H \cdot S^{\dagger} \cdot R_z(\lambda)
\end{equation}
In the case of \lstinline[style=qasm]{U3($\theta$,$\phi$,$\lambda$)}, the equation simplifies as follows.
\begin{equation*}
    U3(\theta, \phi, \lambda) = e^{i\theta} \cdot R_z(\phi) \cdot S \cdot H \cdot R_z(\theta) \cdot H \cdot S^{\dagger} \cdot R_z(\lambda)
\end{equation*}
Note that when $U$ is controlled, if the $S$ and $H$ gates are not controlled, then the $S$ and $H$ gates will cancel out.

\subsection{Decomposing U2 Gates}

The \lstinline[style=qasm]{U2($\phi$,$\lambda$)} gate in OpenQASM is as \lstinline[style=qasm]{gphase(-($\phi$+$\lambda$)/2)} followed by \lstinline[style=qasm]{U($\pi$/2,$\phi$,$\lambda$)}.
Then by Eqn.~\ref{Eqn:UDecomp}, \lstinline[style=qasm]{U2($\phi$,$\lambda$)} has the following decomposition.
\begin{align*}
    U2(\phi, \lambda) &= e^{-i(\phi+\lambda)/2} \cdot U(\pi/2, \phi, \lambda) \\
    &= e^{i(\pi/2)} \cdot R_z(\phi) \cdot S \cdot H \cdot R_z(\pi/2) \cdot H \cdot S^{\dagger} \cdot R_z(\lambda) \\
    &= R_z(\phi) \cdot S \cdot H \cdot \left( e^{i(\pi/2)} \cdot R_z(\pi/2) \right) \cdot H \cdot S^{\dagger} \cdot R_z(\lambda) \\
    &= R_z(\phi) \cdot S \cdot H \cdot P(\pi/2) \cdot H \cdot S^{\dagger} \cdot R_z(\lambda) \\
    &= R_z(\phi) \cdot S \cdot H \cdot S \cdot H \cdot S^{\dagger} \cdot R_z(\lambda)
\end{align*}
Note that when $U2$ is controlled, if the $S \cdot H$ And $H \cdot S^{\dagger}$ subsequences are left uncontrolled, then the $S \cdot H$ and $H \cdot S^{\dagger}$ will cancel out.

\section{Decomposing the W Gate in Quipper}
\label{Appendix:W}

The following decomposition for the $W$ gate is found in~\cite{Crooks2022}.
\begin{center}
    \begin{quantikz}
    & \gate[2]{W} & \\
    & &
\end{quantikz}
=
\begin{quantikz}
    \ghost[2]{W} & \targ{} & \ctrl{1} & \ctrl{1} & \ctrl{1} & \targ{} & \\
    & \ctrl{-1} & \targ{} & \gate{H} & \targ{} & \ctrl{-1} &
\end{quantikz}
\end{center}
Though a proof is never given for this equality, it is easily verified via matrix multiplication.
Note, however, that the controlled $H$ gate is conjugated by controlled $X$ gates with the same control.
Since $X$ is self-inverse, then the controls are unnecessary.
As a result, the following decomposition also holds.
\begin{center}
    \begin{quantikz}
    & \gate[2]{W} & \\
    & &
\end{quantikz}
=
\begin{quantikz}
    \ghost[2]{W} & \targ{} & & \ctrl{1} & & \targ{} & \\
    & \ctrl{-1} & \targ{} & \gate{H} & \targ{} & \ctrl{-1} &
\end{quantikz}
\end{center}

\section{Decomposing the Omega Gate in Quipper}
\label{Appendix:Omega}

The $\omega$ gate in Quipper applies a global phase of $\pi/4$.
However, OpenQASM 2.0 does not offer support for global phase gates.
Instead, we make use of the controlled phase gate $p(\theta)$, which can be implemented in OpenQASM 2.0 up to a global phase.
Note that $X \cdot p(\theta) \cdot X$ implements a negatively controlled phase gate, $X \cdot \diag(1, e^{i\theta}) \cdot X = \diag(e^{i\theta}, 1)$.
Then $P(\theta) \cdot X \cdot P(\theta) X$ applies a global phase of $\theta$.
In particular, $\omega = P(\pi/4) \cdot X \cdot P(\pi/4) \cdot X$.

\section{A Compendium of Gate Decompositions}
\label{Appendix:Decomps}

\begin{figure}[t]
    \begin{center}
        \scalebox{0.97}{\begin{tabular}{@{}ccccccccc@{}}
    \toprule
    Notation & \phantom{a} & Matrix & \phantom{a} & QUIP & \phantom{a} & QASM3 & \phantom{a} & QASM2 \\
  
    \midrule
    \\
    \begin{quantikz} & \targ{} & \end{quantikz}
    && $\begin{bmatrix} 0 & 1 \\ 1 & 0 \end{bmatrix}$
    && $\checkmark$ && $\checkmark$ && $\checkmark$ \\
    \\
    \begin{quantikz} & \gate{Y} & \end{quantikz}
    && $\begin{bmatrix} 0 & -i \\ i & 0 \end{bmatrix}$ &&
    $\checkmark$ && $\checkmark$ && $\checkmark$ \\ 
    \\
    \begin{quantikz} & \gate{Z} & \end{quantikz}
    && $\begin{bmatrix} 1 & 0 \\ 0 & -1 \end{bmatrix}$ &&
    $\checkmark$ && $\checkmark$ && $\checkmark$ \\ 
    \\
    \begin{quantikz} & \gate{H} & \end{quantikz}
    && $\frac{1}{\sqrt{2}} \begin{bmatrix} 1 & 1 \\ 1 & -1 \end{bmatrix}$ &&
    $\checkmark$ && $\checkmark$ && $\checkmark$ \\ 
    \\
    \begin{quantikz} & \gate{S} & \end{quantikz}
    && $\begin{bmatrix} 1 & 0 \\ 0 & i \end{bmatrix}$ &&
    $\checkmark$ && $\checkmark$ && $\checkmark$ \\ 
    \\
    \begin{quantikz} & \gate{S^\dagger} & \end{quantikz}
    && $\begin{bmatrix} 1 & 0 \\ 0 & -i \end{bmatrix}$ &&
    $\checkmark$ && $\checkmark$ && $\checkmark$ \\ 
    \\
    \begin{quantikz} & \gate{T} & \end{quantikz}
    && $\begin{bmatrix} 1 & 0 \\ 0 & \omega \end{bmatrix}$ &&
    $\checkmark$ && $\checkmark$ && $\checkmark$ \\
    \\
    \begin{quantikz} & \gate{T^\dagger} & \end{quantikz}
    && $\begin{bmatrix} 1 & 0 \\ 0 & \omega^7 \end{bmatrix}$ &&
    $\checkmark$ && $\checkmark$ && $\checkmark$ \\
    \\
    \begin{quantikz} & \gate{sx} & \end{quantikz}
    && $\frac{1}{2} \begin{bmatrix} 1+i & 1-i \\ 1-i & 1+i \end{bmatrix}$ &&
    $\checkmark$ && $\checkmark$ && $\times$ \\ 
    \\
    \begin{quantikz} & \gate{iX} & \end{quantikz}
    && $ \begin{bmatrix} 0 & i \\ i & 0 \end{bmatrix}$ &&
    $\checkmark$ && $\times$ && $\times$ \\ 
    \\
    \begin{quantikz} & \gate{\omega} & \end{quantikz}
    && $\begin{bmatrix} \omega & 0 \\ 0 & \omega \end{bmatrix}$ &&
    $\checkmark$ && $\times$ && $\times$ \\ 
    \\
    \begin{quantikz} & \gate{E} & \end{quantikz}
    && $\frac{1}{2} \begin{bmatrix} -1+i & 1+i \\ -1+i & -1-i \end{bmatrix}$ &&
    $\checkmark$ && $\times$ && $\times$ \\
    \\
    \bottomrule
\end{tabular}}
    \end{center}
    \caption{A list of single-qubit non-parametric unitary gates found in either Quipper or OpenQASM.
             Note that $sx$ is called $V$ in Quipper.}
    \label{Tab:1Qubit}
\end{figure}

\begin{figure}[t]
    \begin{center}
        \scalebox{0.94}{\begin{tabular}{@{}ccccccccc@{}}
    \toprule
    Notation & \phantom{a} & Matrix & \phantom{a} & QUIP & \phantom{a} & QASM3 & \phantom{a} & QASM2 \\
  
    \midrule
    \\
    \begin{quantikz} & \ctrl{1} & \\ & \targ{} & \end{quantikz}
    && $I \oplus X$
    && $\checkmark$ && $\checkmark$ && $\checkmark$ \\
    \\
    \begin{quantikz} & \ctrl{1} & \\ & \gate{Y} & \end{quantikz}
    && $I \oplus Y$
    && $\checkmark$ && $\checkmark$ && $\checkmark$ \\
    \\
    \begin{quantikz} & \ctrl{1} & \\ & \phase{} & \end{quantikz}
    && $I \oplus Z$
    && $\checkmark$ && $\checkmark$ && $\checkmark$ \\
    \\
    \begin{quantikz} & \ctrl{1} & \\ & \gate{H} & \end{quantikz}
    && $I \oplus H$
    && $\checkmark$ && $\checkmark$ && $\checkmark$ \\
    \\
    \begin{quantikz} & \gate[2,swap]{} & \\ & & \end{quantikz}
    && $\begin{bmatrix} 1 & 0 & 0 & 0 \\ 0 & 0 & 1 & 0 \\ 0 & 1 & 0 & 0 \\ 0 & 0 & 0 & 1 \end{bmatrix}$
    && $\checkmark$ && $\checkmark$ && $\times$ \\
    \\
    \begin{quantikz} & \gate[2]{W} & \\ & & \end{quantikz}
    && $\begin{bmatrix} 1 & 0 & 0 & 0 \\ 0 & \frac{1}{\sqrt{2}} & \frac{1}{\sqrt{2}} & 0 \\ 0 & \frac{1}{\sqrt{2}} & -\frac{1}{\sqrt{2}} & 0 \\ 0 & 0 & 0 & 1 \end{bmatrix}$
    && $\checkmark$ && $\times$ && $\times$ \\
    \\
    \begin{quantikz} & \ctrl{1} & \\ & \ctrl{1} & \\ & \targ{} & \end{quantikz}
    && $I \oplus I \oplus X$
    && $\checkmark$ && $\checkmark$ && $\checkmark$ \\
    \\
    \begin{quantikz} & \ctrl{1} & \\ & \gate[2,swap]{} & \\ & & \end{quantikz}
    && $I \oplus \mathit{SWAP}$
    && $\checkmark$ && $\checkmark$ && $\times$ \\
    \\
    \bottomrule
\end{tabular}}
    \end{center}
    \caption{A list of multi-qubit non-parametric unitary gates found in either Quipper or OpenQASM.
             Note that in OpenQASM, the controlled $X$, $Y$, $Z$, and $H$ have first-class names.
             For example, a controlled $X$ gate is also a $C(X)$ gate.}
    \label{Tab:MultiQubit}
\end{figure}

\begin{figure}[t]
    \begin{center}
        \scalebox{0.95}{\begin{tabular}{@{}ccccccccc@{}}
    \toprule
    Notation & \phantom{a} & Matrix & \phantom{a} & QUIP & \phantom{a} & QASM3 & \phantom{a} & QASM2 \\
  
    \midrule
    \\
    \begin{quantikz} & \gate{expZ(\theta)} & \end{quantikz}
    && $e^{-iZ\theta}$
    && $\checkmark$ && $\times$ && $\times$ \\
    \\
    \begin{quantikz} & \gate{rGate(\theta)} & \end{quantikz}
    && $\begin{bmatrix} 1 & 0 \\ 0 & e^{2\pi i/\theta} \end{bmatrix}$
    && $\checkmark$ && $\times$ && $\times$ \\
    \\
    \begin{quantikz} & \gate{R_x(\theta)} & \end{quantikz}
    && $e^{-iX(\theta/2)}$
    && $\times$ && $\checkmark$ && $\checkmark$ \\
    \\
    \begin{quantikz} & \gate{R_y(\theta)} & \end{quantikz}
    && $e^{-iY(\theta/2)}$
    && $\times$ && $\checkmark$ && $\checkmark$ \\
    \\
    \begin{quantikz} & \gate{R_z(\theta)} & \end{quantikz}
    && $e^{-iZ(\theta/2)}$
    && $\times$ && $\checkmark$ && $\checkmark$ \\
    \\
    \begin{quantikz} & \gate{P(\theta)} & \end{quantikz}
    && $\begin{bmatrix} 1 & 0 \\ 0 & e^{i\theta} \end{bmatrix}$
    && $\checkmark$ && $\checkmark$ && $\times$ \\
    \\
    \begin{quantikz} & \gate{U(\theta,\phi,\lambda)} & \end{quantikz}
    && See \cref{Appendix:U}.
    && $\times$ && $\checkmark$ && $\checkmark$ \\
    \\
    \begin{quantikz} & \gate{u3(\theta,\phi,\lambda)} & \end{quantikz}
    && See \cref{Appendix:U}.
    && $\times$ && $\checkmark$ && $\checkmark$ \\
    \\
    \begin{quantikz} & \gate{u2(\phi,\lambda)} & \end{quantikz}
    && See \cref{Appendix:U}.
    && $\times$ && $\checkmark$ && $\checkmark$ \\
    \\
    \begin{quantikz} & \gate{u1(\lambda)} & \end{quantikz}
    && See \cref{Appendix:U}.
    && $\times$ && $\checkmark$ && $\checkmark$ \\
    \\
    \bottomrule
\end{tabular}}
    \end{center}
    \caption{A list of single-qubit parametric unitary gates found in either Quipper or OpenQASM.
             Note that $P(\theta)$ is an alias for a controlled global phase of $\theta$, and for the $\mathit{Phase}(\theta)$ gate in OpenQASM.}
    \label{Tab:1Rot}
\end{figure}

\begin{figure}[t]
    \begin{center}
        \scalebox{0.9}{\begin{tabular}{@{}ccccccccc@{}}
    \toprule
    Notation & \phantom{a} & Matrix & \phantom{a} & QUIP & \phantom{a} & QASM3 & \phantom{a} & QASM2 \\
  
    \midrule
    \\
    \begin{quantikz} & \ctrl{1} & \\ & \gate{R_x(\theta)} & \end{quantikz}
    && $I \oplus R_x(\theta)$
    && $\times$ && $\checkmark$ && $\times$ \\
    \\
    \begin{quantikz} & \ctrl{1} & \\ & \gate{R_y(\theta)} & \end{quantikz}
    && $I \oplus R_y(\theta)$
    && $\times$ && $\checkmark$ && $\times$ \\
    \\
    \begin{quantikz} & \ctrl{1} & \\ & \gate{R_z(\theta)} & \end{quantikz}
    && $I \oplus R_z(\theta)$
    && $\times$ && $\checkmark$ && $\checkmark$ \\
    \\
    \begin{quantikz} & \ctrl{1} & \\ & \gate{P(\theta)} & \end{quantikz}
    && $I \oplus P(\theta)$
    && $\times $ && $\checkmark$ && $\times$ \\
    \\
    \begin{quantikz} & \gate[2]{CU(\gamma, \theta, \rho, \lambda)} & \\ & & \end{quantikz}
    && $e^{i\gamma} \cdot U(\theta, \rho, \lambda)$
    && $\times$ && $\checkmark$ && $\times$ \\
    \\
    \bottomrule
\end{tabular}}
    \end{center}
    \caption{A list of multi-qubit parametric unitary gates found in either Quipper or OpenQASM.
             Note that in OpenQASM, the controlled $R_x$, $R_y$, $R_z$, and $P$ have first-class names.
             For example, a controlled $R_x$ gate is also a $CR_x$ gate.
             However, the $CU(\gamma, \theta, \phi, \lambda)$ gate differs from the controlled $U(\theta, \phi, \lambda)$, as suggested by the additional parameter.
             Note also that $CP(\theta)$ is an alias for a doubly-controlled global phase of $\theta$, and for the $\mathit{CPhase}(\theta)$ gate in OpenQASM.}
    \label{Tab:MultiRot}
\end{figure}

\cref{Tab:1Qubit} provides a list of single-qubit non-parametric unitary gates, their matrix semantics, and the languages for which the gate is supported.
Likewise, multi-qubit non-parametric unitary gates are found in \cref{Tab:MultiQubit}, single-qubit parameteric unitary gates are found in \cref{Tab:1Rot}, and multi-qubit parametric unitary gates are found in \cref{Tab:MultiRot}.
LinguaQuanta provides translations for all gates listed above.
This section outlines all necessary rewrite rules, together with the decompositions for inverse and control modifiers.
Decompositions taken from the literature are cited accordingly.
Several decompositions are novel and are found in \cref{Appendix:U,Appendix:W,Appendix:Omega}.
All other decompositions were either trivial, or obtained through trial-and-error with validation via matrix multiplication.

\subsection{Translating Quipper to OpenQASM}
\label{Appendix:Decomps:QuipToQasm}

The following decompositions are found in the \texttt{quipgates.inc} runtime library.
\begin{align*}
    \begin{quantikz}
    & \gate{iX} &
\end{quantikz}
&=
\begin{quantikz}
    & \targ{} & \gate{S} & \targ{} & \gate{S} & \targ{} &
\end{quantikz}
    \\
    \begin{quantikz}
    & \gate{\omega} &
\end{quantikz}
&=
\begin{quantikz}
    & \gate{P \left( \frac{\pi}{4} \right)} & \targ{} & \gate{P \left( -\frac{\pi}{4} \right)} & \targ{} &
\end{quantikz}
    \\
    \begin{quantikz}
    & \gate{E} &
\end{quantikz}
&=
\begin{quantikz}
    & \gate{\omega} & \gate{\omega} & \gate{\omega} & \gate{S} & \gate{S} & \gate{S} & \gate{H} &
\end{quantikz} \text{\;\cite{Selinger2014}}
    \\
    \begin{quantikz}
    & \gate[2]{W} & \\
    & &
\end{quantikz}
&=
\begin{quantikz}
    \ghost[2]{W} & \targ{} & & \ctrl{1} & & \targ{} & \\
    & \ctrl{-1} & \targ{} & \gate{H} & \targ{} & \ctrl{-1} &
\end{quantikz}
    \\
    \begin{quantikz}
    & \gate{expZ(\theta)} &
\end{quantikz}
&=
\begin{quantikz}
    & \gate{R_z(2 \cdot \theta)} &
\end{quantikz}
    \\
    \begin{quantikz}
    & \gate{rGate(\theta)} &
\end{quantikz}
&=
\begin{quantikz}
    & \gate{P \left( \frac{2 \pi i}{\theta} \right)} &
\end{quantikz}
\end{align*}

\subsection{Translating OpenQASM to Quipper}

This section provides Quipper decompositions for the OpenQASM-specific gates found in \cref{Tab:1Rot,Tab:MultiRot}.
Rotations are decomposed using Decomposition~1 in \cref{Sect:Decompositions}, together with the relations that $HZH = X$ and $(XS^{\dagger}H)Z(HSX) = Y$.
Controls are optimized by observing that conjugate gates cancel out when left uncontrolled.
The $U$ gates are decomposed as described in \cref{Sect:UGate}.

\begin{align*}
    \begin{quantikz}
    & \gate{Rx(\theta)} &
\end{quantikz}
&=
\begin{quantikz}
    & \gate{H} & \gate{expZ \left( \frac{\theta}{2} \right)} & \gate{H} &
\end{quantikz}
    \\
    \begin{quantikz}
    & \ctrl{1} & \\
    & \gate{Rx(\theta)} &
\end{quantikz}
&=
\begin{quantikz}
    & & \ctrl{1} & & \\
    & \gate{H} &  \gate{expZ \left( \frac{\theta}{2} \right)} & \gate{H} &
\end{quantikz}
    \\
    \begin{quantikz}
    & \gate{Ry(\theta)} &
\end{quantikz}
&=
\begin{quantikz}
    & \targ{} & \gate{S} & \gate{H} &  \gate{expZ \left( \frac{\theta}{2} \right)} & \gate{H} & \gate{S^\dagger} & \targ{} &
\end{quantikz}
    \\ 
    \begin{quantikz}
    & \ctrl{1} & \\
    & \gate{Ry(\theta)} &
\end{quantikz}
&=
\begin{quantikz}
    & & & & \ctrl{1} & & & & \\
    & \targ{} & \gate{S} & \gate{H} &  \gate{expZ \left( \frac{\theta}{2} \right)} & \gate{H} & \gate{S^\dagger} & \targ{} &
\end{quantikz} 
    \\
    \begin{quantikz}
    & \gate{Rz(\theta)} &
\end{quantikz}
&=
\begin{quantikz}
    &\gate{expZ \left( \frac{\theta}{2} \right)} &
\end{quantikz}
    \\
    \begin{quantikz}
    & \gate{u1(\lambda)} &
\end{quantikz}
&=
\begin{quantikz}
    & \phase{\lambda/2} & \gate{expZ \left( \frac{\lambda}{2} \right)} &
\end{quantikz}
\end{align*}
For decompositions that do not fit inline, see \cref{Fig:U2,Fig:CtrlU2,Fig:U3,Fig:CtrlU3,Fig:U,Fig:CtrlU,Fig:CU}.

\subsection{Translating OpenQASM 3 to OpenQASM 2.0}

The following decompositions are found in the \texttt{bkpgates.inc} runtime library.
Note that the $R_z(\theta)$ gate in OpenQASM 2.0 is equivalent to the $P(\theta)$ gate in OpenQASM 3, due to differences in global phase~\cite{CrossJavadiAbhari2022}.
All controlled rotations are decomposed using Decomposition~2 in \cref{Sect:Decompositions}.
The $P(\theta)$ gate is decomposed using Decomposition~4 in \cref{Sect:Decompositions}.
Note that $HXH = Z$, $Z \cdot R_x(\theta) \cdot Z = R_x(-\theta)$, and $Z \cdot R_y(\theta) \cdot Z = R_y(-\theta)$.
\begin{align*}
    \begin{quantikz}
    & \gate[2,swap]{} & \\
    & &
\end{quantikz}
&=
\begin{quantikz}[row sep = 0.65cm]
    & \ctrl{1} & \targ{} & \ctrl{1} & \\
    & \targ{} & \ctrl{-1} & \targ{} &
\end{quantikz}
    \\
    \begin{quantikz}
    & \ctrl{1} & \\
    & \gate[2,swap]{} & \\
    & &
\end{quantikz}
&=
\begin{quantikz}[row sep = 0.65cm]
    & & \ctrl{1} & & \\
    & \ctrl{1} & \targ{} & \ctrl{1} & \\
    & \targ{} & \ctrl{-1} & \targ{} &
\end{quantikz} \text{\;\cite{NielsenChuang2011}}
    \\
    \begin{quantikz}
    & \gate{sx} &
\end{quantikz}
&=
\begin{quantikz}
    & \gate{H} & \gate{S^\dagger} & \gate{H} &
\end{quantikz} \text{\;\cite{Selinger2014}}
    \\
    \begin{quantikz}
    & \ctrl{1} & \\
    & \gate{R_x(\theta)} &
\end{quantikz}
&=
\begin{quantikz}
    & & \ctrl{1} & & \ctrl{1} & \\
    & \gate{R_x\left(\frac{\theta}{2}\right)} & \phase{} & \gate{R_x\left(-\frac{\theta}{2}\right)} & \phase{} &
\end{quantikz}
    \\
    \begin{quantikz}
    & \ctrl{1} & \\
    & \gate{R_y(\theta)} &
\end{quantikz}
&=
\begin{quantikz}
    & & \ctrl{1} & & \ctrl{1} & \\
    & \gate{R_y\left(\frac{\theta}{2}\right)} & \phase{} & \gate{R_y\left(-\frac{\theta}{2}\right)} & \phase{} &
\end{quantikz}
    \\
    \begin{quantikz}
    \ghost{P\left(\frac{\theta}{2}\right)} & \ctrl{1} & \\
    & \gate{P(\theta)} &
\end{quantikz}
&=
\begin{quantikz}
    & & \ctrl{1} & & \ctrl{1} & \gate{P\left(\frac{\theta}{2}\right)} & \\
    & \gate{P\left(\frac{\theta}{2}\right)} & \targ{} & \gate{P\left(-\frac{\theta}{2}\right)} & \targ{} & &
\end{quantikz}
\end{align*}

\subsection{Decomposing Inverse Modifiers}

The following decompositions are used to eliminate inverse modifiers.
Note that the gates $X$, $Y$, $Z$, $H$, $\mathit{SWAP}$, $W$, $C(X)$, $C(Y)$, $C(Z)$, $C(H)$, $C(\mathit{SWAP})$, and $C(C(X))$ are all self-inverse, and are therefore omitted.
Note that the gates $S$ and $T$ are omitted since $S^{\dagger}$ and $T^{\dagger}$ already exist in OpenQASM.
 \begin{align*}
    \begin{quantikz}
    & \gate{R_x(\theta)^{\dagger}} &
\end{quantikz}
&=
\begin{quantikz}
    & \gate{R_x(-\theta)} &
\end{quantikz}
    &
    \begin{quantikz}
    & \gate{R_y(\theta)^{\dagger}} &
\end{quantikz}
&=
\begin{quantikz}
    & \gate{R_y(-\theta)} &
\end{quantikz}
    \\
    \begin{quantikz}
    & \gate{R_z(\theta)^{\dagger}} &
\end{quantikz}
&=
\begin{quantikz}
    & \gate{R_z(-\theta)} &
\end{quantikz}
    &
    \begin{quantikz}
    & \gate{P(\theta)^{\dagger}} &
\end{quantikz}
&=
\begin{quantikz}
    & \gate{P(-\theta)} &
\end{quantikz}
    \\
    \begin{quantikz}
    & \ctrl{1} & \\
    & \gate{R_x(\theta)^{\dagger}} &
\end{quantikz}
&=
\begin{quantikz}
    & \ctrl{1} & \\
    & \gate{R_x(-\theta)} &
\end{quantikz}
    &
    \begin{quantikz}
    & \ctrl{1} & \\
    & \gate{R_y(\theta)^{\dagger}} &
\end{quantikz}
&=
\begin{quantikz}
    & \ctrl{1} & \\
    & \gate{R_y(-\theta)} &
\end{quantikz}
    \\
    \begin{quantikz}
    & \ctrl{1} & \\
    & \gate{R_z(\theta)^{\dagger}} &
\end{quantikz}
&=
\begin{quantikz}
    & \ctrl{1} & \\
    & \gate{R_z(-\theta)} &
\end{quantikz}
    &
    \begin{quantikz}
    & \ctrl{1} & \\
    & \gate{P(\theta)^{\dagger}} &
\end{quantikz}
&=
\begin{quantikz}
    & \ctrl{1} & \\
    & \gate{P(-\theta)} &
\end{quantikz}
\end{align*}
\begin{align*}
    \begin{quantikz}
    & \gate{\omega^\dagger} &
\end{quantikz}
&=
\begin{quantikz}
    \ghost{\omega^\dagger} & \phase{7\frac{\pi}{4}} &
\end{quantikz}
    &
    \begin{quantikz}
    & \gate{iX^\dagger} &
\end{quantikz}
&=
\begin{quantikz}
    & \phase{\pi} & \gate{iX} &
\end{quantikz}
    \\
    \begin{quantikz}
    & \gate{sx^\dagger} &
\end{quantikz}
&=
\begin{quantikz}
    & \gate{sx} & \targ{} &
\end{quantikz}
    &
    \begin{quantikz}
    & \gate{E^\dagger} &
\end{quantikz}
&=
\begin{quantikz}
    & \phase{5\frac{\pi}{4}} & \gate{H} & \gate{S} &
\end{quantikz}
\end{align*}
\begin{align*}
    \begin{quantikz}
    & \gate{U(\theta, \phi, \lambda)^\dagger} &
\end{quantikz}
&=
\begin{quantikz}
    & \gate{U(-\theta, -\lambda, -\phi)} &
\end{quantikz} \text{\;\cite{CrossJavadiAbhari2022}}
    \\
    \begin{quantikz}
    & \gate{u3(\theta, \phi, \lambda)^\dagger} &
\end{quantikz}
&=
\begin{quantikz}
    & \gate{u3(-\theta, -\lambda, -\phi)} &
\end{quantikz}
    \\
    \begin{quantikz}
    & \gate{u2(\phi, \lambda)^\dagger} &
\end{quantikz}
&=
\begin{quantikz}
    & \gate{u3 \left( -\frac{\pi}{2}, -\lambda, -\phi \right)} &
\end{quantikz}
    \\
    \begin{quantikz}
    & \gate{u1(\lambda)^\dagger} &
\end{quantikz}
&=
\begin{quantikz}
    & \gate{u1(-\lambda)} &
\end{quantikz}
    \\
    \begin{quantikz}
    & \gate[2]{CU(\theta, \phi, \lambda, \gamma)^\dagger} & \\
    & &
\end{quantikz}
&=
\begin{quantikz}
    & \gate[2]{CU(-\theta, -\lambda, -\phi, -\gamma)} & \\
    & &
\end{quantikz} \text{\;\cite{CrossJavadiAbhari2022}}
\end{align*}

\subsection{Decomposing Control Modifiers}

The following decompositions are used to eliminate controls from Quipper gates.
Explicit decompositions are given for all gates with single controls, and several gates with two controls.
For arbitrary rotations, if the end-user provides a unitary $D$, then Decomposition~3 is used from \cref{Sect:Decompositions}.
To deal with additional controls, Decomposition~2 is used from \cref{Sect:Decompositions}.
The gates $C(X)$, $C(Y)$, $C(Z)$, $C(H)$, $C(\mathit{SWAP})$, $P(\theta)$, and $C(P(\theta))$ are used as base cases without decomposition, since they are supported in OpenQASM 3.
\begin{align*}
    \begin{quantikz}
    \ghost{P\left( \frac{\pi}{4} \right)} & \ctrl{1} & \\
    & \gate{\omega} &
\end{quantikz}
&=
\begin{quantikz}
    & \gate{P\left( \frac{\pi}{4} \right)} & \\
    & \ghost{\omega} &
\end{quantikz}
    \\
    \begin{quantikz}
    \ghost{S} & \ctrl{1} & \\
    & \gate{iX} &
\end{quantikz}
&=
\begin{quantikz}
    & \ctrl{1} & \gate{S} & \\
    \ghost{iX} & \targ{} & &
\end{quantikz}
    \\
    \begin{quantikz}
    \ghost{T^\dagger} & \ctrl{1} & \\
    & \gate{S} &
\end{quantikz}
&=
\begin{quantikz}
    & \targ{} & \gate{T^\dagger} & \targ{} & \gate{T} & \\
    & \ctrl{0}\wire[u]{q} & & \ctrl{0}\wire[u]{q} & \gate{T} &
\end{quantikz} \text{\;\cite{AmyMaslov2013}}
    \\
    \begin{quantikz}
    \ghost{T^\dagger} & \ctrl{1} & \\
    \ghost{T^\dagger} & \gate{V} &
\end{quantikz}
&=
\begin{quantikz}
    & \gate{T^\dagger} & \targ{} & \gate{T} & \targ{} & & \\
    & \gate{H} & \ctrl{0}\wire[u]{q} & \gate{T^\dagger} & \ctrl{0}\wire[u]{q} & \gate{H} &
\end{quantikz} \text{\;\cite{AmyMaslov2013}}
    \\
    \begin{quantikz}
    \ghost{S} & \ctrl{1} & \\
    \ghost{T^\dagger} & \gate{E} &
\end{quantikz}
&=
\begin{quantikz}
    & \gate{S} & & \ctrl{1} & & & \targ{} & \gate{T} & \targ{} & \\
    & \gate{H} & \gate{T} & \targ{} & \gate{T^\dagger} & \gate{H} & \ctrl{0}\wire[u]{q} & \gate{T^\dagger} & \ctrl{0}\wire[u]{q} &
\end{quantikz} \text{\;\cite{Selinger2014}}
\end{align*}
For decompositions that do not fit inline, see \cref{Fig:W,Fig:T,Fig:CCZ}.

\subsection{Control Decompositions for OpenQASM 2.0}

The following decompositions are used to eliminate controlled gates without legacy support.
\begin{align*}
    \begin{quantikz}
    & \ctrl{1} & \\
    & \gate{H} &
\end{quantikz}
&=
\begin{quantikz}
    & & & & \ctrl{1} & & & & \\
    & \gate{S} & \gate{H} & \gate{T} & \targ{} & \gate{T^\dagger} & \gate{H} & \gate{S^\dagger} &
\end{quantikz} \text{\;\cite{AmyMaslov2013}}
\end{align*}
For decompositions that do not fit inline, see \cref{Fig:CCiX,Fig:Toffoli,Fig:Swap}.

\begin{landscape}
    \hspace{0pt}
    \vfill

    \begin{figure}
        \centering
        \input{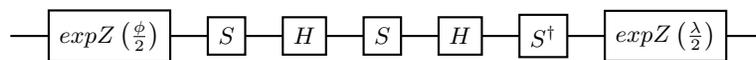}
        \caption{Decomposition of a $u2(\phi, \lambda)$ gate.}
        \label{Fig:U2}
    \end{figure}

    \begin{figure}
        \centering
        \input{circuits/appendix/rot/decomp_ctrl_u2}
        \caption{Decomposition of an $n$-fold controlled $u2(\phi, \lambda)$ gate. Illustrated for $n=1$.}
        \label{Fig:CtrlU2}
    \end{figure}

    \vfill
    \hspace{0pt}
\end{landscape}

\begin{landscape}
    \hspace{0pt}
    \vfill

    \begin{figure}[!htb]
        \centering
        \scalebox{0.9}{\input{circuits/appendix/rot/decomp_u3}}
        \caption{Decomposition of a $u3(\theta, \phi, \lambda)$ gate.}
        \label{Fig:U3}
    \end{figure}

    \begin{figure}[!htb]
        \centering
        \scalebox{0.9}{\input{circuits/appendix/rot/decomp_ctrl_u3}}
        \caption{Decomposition of an $n$-fold controlled $u3(\theta, \phi, \lambda)$ gate. Illustrated for $n=1$.}
        \label{Fig:CtrlU3}
    \end{figure}

    \begin{figure}[!htb]
        \centering
        \scalebox{0.9}{\input{circuits/appendix/rot/decomp_u}}
        \caption{Decomposition of a $U(\theta, \phi, \lambda)$ gate.}
        \label{Fig:U}
    \end{figure}

    \vfill
    \hspace{0pt}
\end{landscape}

\begin{landscape}
    \hspace{0pt}
    \vfill

    \begin{figure}
        \centering
        \scalebox{0.9}{\input{circuits/appendix/rot/decomp_ctrl_u}}
        \caption{Decomposition of an $n$-fold controlled $U(\theta, \phi, \lambda)$ gate. Illustrated for $n=1$.}
        \label{Fig:CtrlU}
    \end{figure}

    \begin{figure}
        \centering
        \scalebox{0.9}{\input{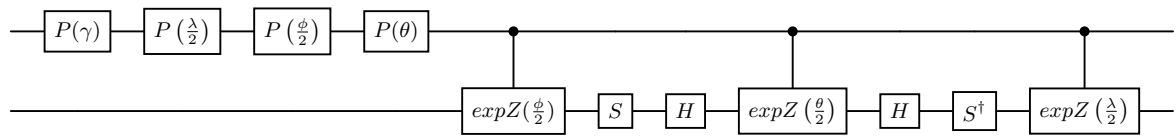}}
        \caption{Decomposition of a $CU(\theta, \phi, \lambda, \gamma)$ gate. Note that this differs from a controlled $U$ gate.}
        \label{Fig:CU}
    \end{figure}

    \vfill
    \hspace{0pt}
\end{landscape}

\begin{landscape}
    \hspace{0pt}
    \vfill

    \begin{figure}
        \centering
        \scalebox{0.9}{\input{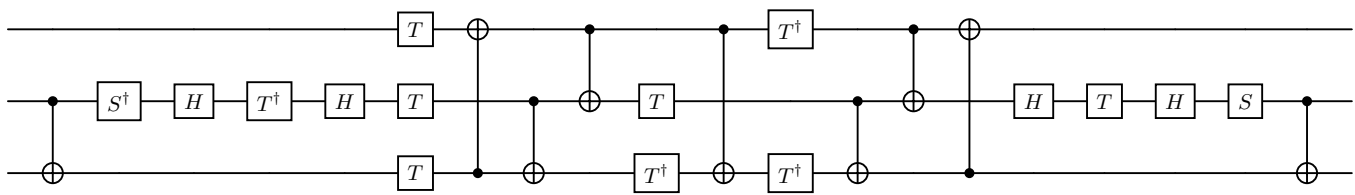}}
        \caption{Decomposition of a controlled $W$ gate from~\cite{Selinger2014} based on results from~\cite{AmyMaslov2013}.}
        \label{Fig:W}
    \end{figure}

    \begin{figure}
        \centering
        \scalebox{0.9}{\input{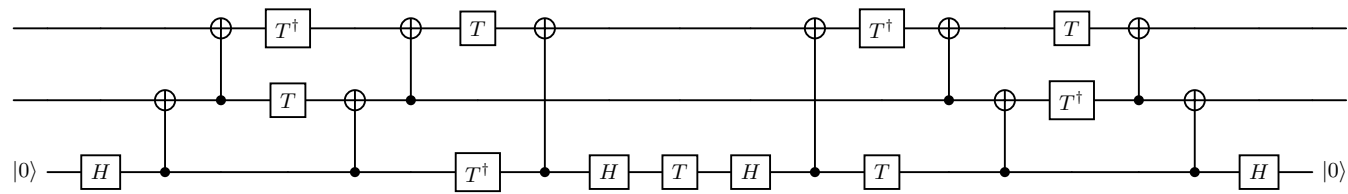}}
        \caption{Decomposition of a controlled $T$ gate~\cite{AmyMaslov2013}.}
        \label{Fig:T}
    \end{figure}

    \vfill
    \hspace{0pt}
\end{landscape}

\begin{landscape}
    \hspace{0pt}
    \vfill
    
    \begin{figure}
        \centering
        \input{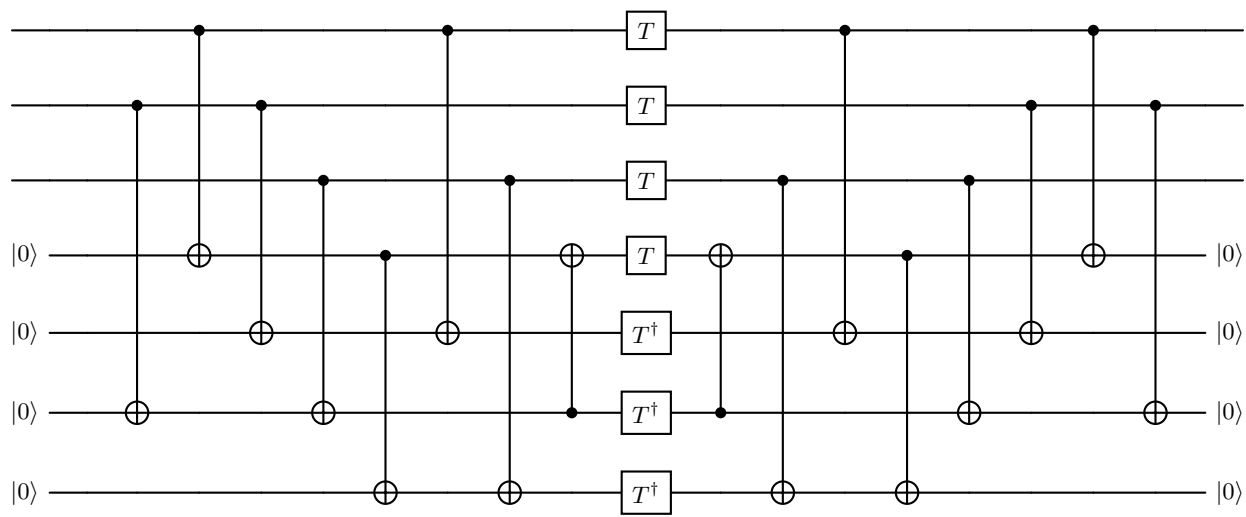}
        \caption{Decomposition of a doubly-controlled $Z$ gate~\cite{Selinger2013}.}
        \label{Fig:CCZ}
    \end{figure}

    \vfill
    \hspace{0pt}
\end{landscape}

\begin{landscape}
    \hspace{0pt}
    \vfill
    
    \begin{figure}
        \centering
        \input{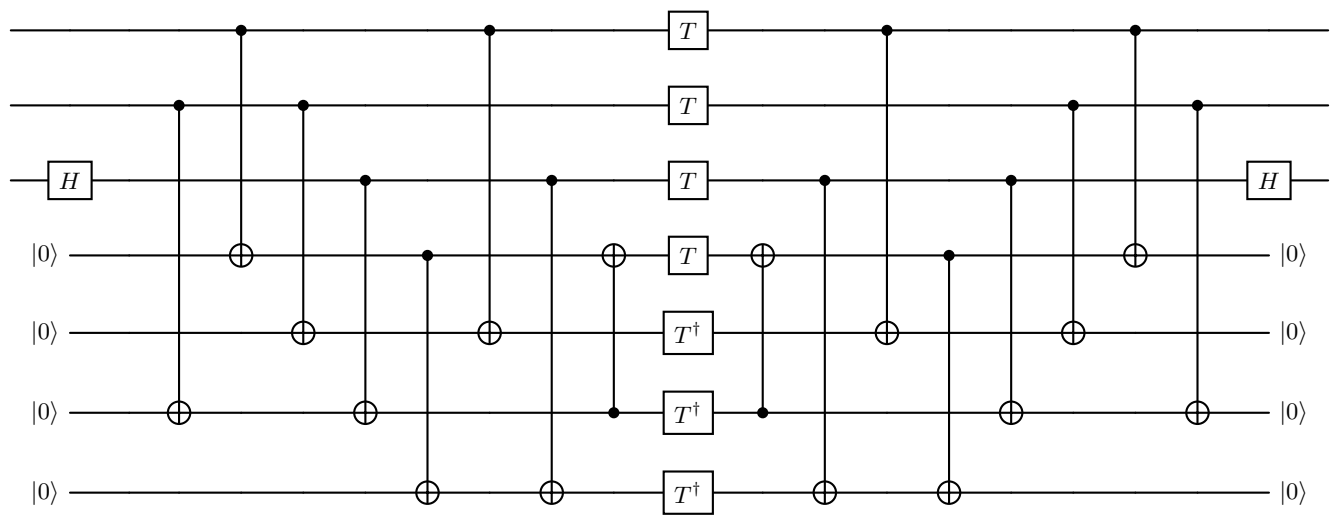}
        \caption{Decomposition of a Toffoli gate~\cite{Selinger2013}.}
        \label{Fig:Toffoli}
    \end{figure}

    \vfill
    \hspace{0pt}
\end{landscape}

\begin{landscape}
    \hspace{0pt}
    \vfill

    \begin{figure}
        \centering
        \scalebox{0.8}{\input{circuits/appendix/ctrl/decomp_ccix}}
        \caption{Decomposition of a doubly-controlled $iX$ gate~\cite{GilesSelinger2013}.}
        \label{Fig:CCiX}
    \end{figure}

    \begin{figure}
        \centering
        \scalebox{0.8}{\input{circuits/appendix/ctrl/decomp_swap}}
        \caption{Decomposition of a controlled swap gate~\cite{AmyMaslov2013}.}
        \label{Fig:Swap}
    \end{figure}

    \vfill
    \hspace{0pt}
\end{landscape}

}

\end{document}